\documentclass[twocolumn]{aastex62}
\usepackage{amsmath}
\usepackage{natbib}
\usepackage{outlines}

\graphicspath{{./}{figures/}}

\received{\today}
\revised{TBD}
\accepted{TBD}
\submitjournal{ApJ}

\shorttitle{Missing Planets}
\shortauthors{Brittain et al.}

\begin{document}

\title{The Planetary Luminosity Problem: ``Missing Planets'' and the Observational Consequences of Episodic Accretion}

\correspondingauthor{Sean D. Brittain}
\email{sbritt@clemson.edu}

\author[0000-0001-5638-1330]{Sean D. Brittain}
\altaffiliation{Visiting Scientist, NOAO}
\affiliation{Clemson University \\
118 Kinard Laboratory
Clemson, SC 29634, USA}

\author{Joan R. Najita}
\affiliation{National Optical Astronomical Observatory \\
950 North Cherry Avenue \\ 
Tucson, AZ 85719, USA}

\author[0000-0001-9290-7846]{Ruobing Dong}
\affiliation{Department of Physics \& Astronomy \\
University of Victoria \\
Victoria BC V8P 1A1, Canada}

\author[0000-0003-3616-6822]{Zhaohuan Zhu}
\affiliation{Department of Physics and Astronomy \\
University of Nevada, Las Vegas\\
4505 S. Maryland Pkwy. \\
Las Vegas, NV 89154, USA }

\begin{abstract}

The high occurrence rates of spiral arms and large central clearings in 
protoplanetary disks, if interpreted as signposts of giant planets, 
indicate that gas giants form commonly as companions to 
young stars ($<$ few Myr) at orbital separations of 10--300\,au. 
However, attempts to directly image this giant planet population as 
companions to more mature stars ($> 10$\, Myr) have yielded few successes. 
This discrepancy could be explained if most giant planets form 
``cold start,'' i.e., by radiating away much of their formation energy 
as they assemble their mass, rendering them faint enough to elude 
detection at later times. In that case, giant planets should be bright
at early times, during their accretion phase, and yet forming 
planets are detected only rarely through direct imaging techniques. 
Here we explore the possibility that the low detection rate of accreting 
planets is the result of episodic accretion through a circumplanetary 
disk. We also explore the 
possibility that the companion orbiting the Herbig Ae star HD~142527 
may be a giant planet undergoing such an accretion outburst.

\end{abstract}


\section{Introduction} 
Giant planets are common at small orbital separations, 
as are indirect signatures of their existence 
at large separations. 
At orbital separations $\lesssim 10$\,au, 10-20\% of low mass stars ($< 1.5$ $M_\odot$; LMS), 
host a gas giant planet ($\gtrsim 1\,M_J;$ \citealt{Cassan2012, Cumming2008}).
At large separations ($\gtrsim 10$\,au), 
a high occurrence rate of giant planetary companions 
to LMS is also inferred, based on the morphologies 
of protoplanetary disks surrounding young LMS (i.e., T Tauri stars). 
Some 10-15\% of T Tauri disks have 
the spectral energy distribution (SED) 
of a transition 
disk, i.e., a protoplanetary disk in which the central portion  
($\lesssim$10--40\,au) is optically thin in the infrared continuum 
\citep{muz2010}. 
%a configuration that has been confirmed by 
%submillimeter continuum imaging \citep{Andrews2011}.
Transition disk morphologies and other corroborating properties are 
interpreted as the signpost of one or more giant planets 
($\gtrsim 3\,M_J)$ orbiting within the optically thin region of the disk 
\citep{dodson2011, zhu2011, Espaillat2014}.

A similar story is found for giant planet companions to 
intermediate mass stars ($\sim$1.5--2\,M$_{\sun}$; IMS). 
At small orbital separations, radial velocity studies 
find that $\sim$14\% of IMS harbor a 
gas giant planet within $\sim$10~au of the star 
\citep{Luhn2018,Johnson2010}. 
Transition disk SEDs are also common among young IMS (i.e., Herbig Ae stars); 
approximately 40\% of Herbig Ae stars within 200 pc 
are transition disk sources \citet{Brittain2019b}.
In addition, a surprisingly large fraction of 
well-studied Herbig Ae disks show dramatic two-arm spirals 
($\sim$20\,\% \citealt{dong2018a}), which point to the presence of 
high mass giant planets (5--13$\,M_J$) at 30--300\,au \citep{Fung2015spiral, Dong2017spiral}.

In contrast to the high frequency of indirect indicators 
of giant planets beyond $\sim$10\,au, direct detection of the planets 
themselves has proven challenging as companions to both mature stars 
as well as young stars surrounded by protoplanetary disks. 
Many high contrast imaging surveys have searched for 
giant planet companions to stars older than 10~Myr 
at orbital separations $>$10\,au. 
As summarized by 
\citet[see also \citealt{Nielsen2019}]{Bowler2016},  %\rdnote{Nielsen+19 actually reported a somewhat different number; %2.8\% only comes from the Bowler 2016 paper }, 
such studies 
find that high mass planets (5--13\,$M_J$) 
are detected at 30--300\,au orbital separation 
in only a small fraction of mature IMS \citep[2.8\%][]{Bowler2016}, 
assuming they are as bright 
as predicted by ``hot start'' planetary evolutionary models. 
The low incidence rate is much less than the $\sim$20\% 
two-arm spiral arm fraction of young IMS that points to planets in the same range of mass and orbital separation
\citep{dong2018a}. 
For mature LMS stars, 
%fewer than 7\% have a 5--10\,$M_J$ planet at 30--300\,au \citet{Bowler2016} and
$\sim 1$\% have a 1--13\,$M_J$ giant planet at 10--100\,au 
(Nielsen et al.\ 2019, Figure 18), a much lower occurrence 
rate than the 10--15\% occurrence rate of transition disks among the T Tauri star 
population. 

The discrepancy between the detection rates of indirect signposts
of planet formation and of the planets themselves 
could indicate
that the indirect signposts are caused by something other than planets.
For example, two-armed spirals can also 
arise from gravitationally unstable disks \citep{Dong2015GI, Kratter2016}. However,
the lifetime of the gravitationally unstable 
phase is too short to account for the large number of two-armed 
spirals observed around young stars \citep[][]{dong2018a, hall2018}.

Alternatively, the discrepantly low detection rate of high mass giant planets as companions 
to mature stars could be entirely due to the use of the hot start models, 
which assume that planets retain their heat of formation 
(i.e., gravitational potential energy) when they form. 
The alternative ``cold start'' models \citep{marley2007, fortney2008} assume that 
planets radiate away much of their accretion energy in the formation phase. 
As a result, they predict considerably fainter planets at ages 
$>$10\,Myr compared to hot start models. 
\citet{Stone2018} found that essentially all nearby FGK stars 
could harbor one or more 7--10\,$M_J$ planets at 5--50\,au if they formed 
cold start. The planets are simply too faint to be detected with 
current surveys. 

While the above discrepancy 
could point to the validity of 
cold start models over hot start models, that scenario  
implies that accreting planets would be very bright 
in their youth as they radiate away their accretion energy 
in accretion shocks and/or through a circumplanetary disk. 
That is, young giant planets ($<$ few Myr) embedded in 
protoplanetary disks should be readily detectable  
during their runaway accretion phase \citep{Eisner2015, Zhu2015}. 

So it is perhaps surprising that giant planets are detected 
infrequently as companions to young stars surrounded by 
protoplanetary disks. 
While transition disks have been targeted in many high 
contrast imaging studies (e.g., Subaru SEEDS), 
planetary companion candidates have been detected in only a few sources. 
The best candidates to date are the companions to two 
LMS transition disks---PDS~70 \citep{Keppler2018} and LkCa~15 \citep{Kraus2012, Sallum2015}---and the companions to the IMS 
transition disk HD~100546 
\citep{Perez2019, Brittain2019}. 
As in the case of the spatially resolved (stellar) companion 
to the IMS transition disk HD~142527 \citep{Biller2012, Close2014}, 
H$\alpha$ emission from the PDS~70 and LkCa~15 companions suggests that they are actively 
accreting.

Here we explore the reason for the infrequent 
detection of accreting giant planets in protoplanetary disks 
despite the likelihood that such planets occur commonly and 
radiate away their accretion energy as they accrete.
We propose that the low ``luminosity problem'' of forming planets  
is solved in a similar fashion to the luminosity problem of foming stars
\citep{kenyon1990, Kenyon1995}. Namely, forming planets accrete episodically, 
much like how forming stars undergo
%in a similar way to forming stars undergoing 
FU~Ori-like outbursts. 
Previous studies have argued that circumplanetary disks, like 
circumstellar disks, are likely to harbor significant dead zones 
and would accrete episodically \citep{Lubow2012}, or that vortices form 
in circumplanetary disks and generate short-timescale outbursts \citep{Zhu2016}.

To explore this scenario in the context of the developing 
detection statistics and properties of young planetary companions, 
we summarize in section 2 the properties of resolved planetary 
companions to young protoplanetary disk sources. 
In section 3, we describe a simple toy model for episodic accretion 
based on the theoretical literature and illustrate how 
episodic accretion can potentially account for the low detection 
rate of accreting planets. 
In section 4, we explore the possibility that the companion to 
HD~142527 is actually a planetary companion undergoing an accretion 
outburst rather than a low mass stellar companion and discuss 
ways to discriminate between the two possibilities. 
We conclude with a discussion of our results and 
opportunities for future progress.

\section{Searches for Young Planets}

The largest published near infrared imaging survey for planets 
forming in protoplanetary disks was carried out 
in the SEEDS campaign using Subaru/HiCIAO. \citet{Uyama2017} reported 
SEEDS observations of 68 YSOs (39 LMS and 29 IMS) at NIR wavelengths. Some of 
the targets were primarily observed in polarization differential 
imaging mode to probe disk structures, while a fraction of the 
targets were observed in angular differential imaging (ADI) mode 
under decent conditions. Specifically, for 20 
targets a 5-$\sigma$ contrast level of 5 magnitudes or more was 
achieved at $H$-band at 0$\arcsec$.25 separation, which corresponds
to a typical upper limit on a planet's $H$-band luminosity of $3\times10^{31}$~erg~s$^{-1}~$\micron$^{-1}$ (Table 1)
and a planet mass limit of 5-10 M$\rm_J$ at a few tens of AU, 
assuming hot start models \citep{Baraffe2003}. 
These results are 
based on conventional ADI data reduction, and have not taken into 
account the effects of circumplanetary material on the detectability 
of embedded planets; \citealt{maire2017}). 

Other searches using ADI, carried out for planets forming in individual 
protoplanetary disks, have probed smaller orbital separations
$\sim 0.1\arcsec$
and lower companion luminosities
\citep[e.g.,][]{ quanz2013, Cieza2013,  Currie2015, 
 Testi2015,  canovas2017, currie2017, 
Follette2017, maire2017, Guidi2018, Keppler2018, 
Ligi2018, Reggiani2018k, Sissa2018, cugno2019, Gratton2019}. Similarly, aperture masking
has been used to probe even smaller separations \citep[e.g.,][]{Biller2012, 
Kraus2012,Grady2013,Kraus2013,Sallum2015,Willson2016}. The results of these observations are summarized in Table 1. 

In addition to NIR searches for forming planets in disks, there have been several efforts to image H$\alpha$ emission arising
from the accretion shock on the forming planet. Several studies have targeted individual objects (HD~142527 -- \citealt{Close2014}; LkCa~15 -- \citealt{Sallum2015}; PDS70 -- \citealt{haffert2019, Wagner2018}). In a recent study, \citet{zurlo2020} studied 11 nearby transitions disks with SPHERE on the VLT and found no H$\alpha$ emission from accreting planets down to an upper limit on accretion luminosity of 10$^{-6}$~L$_{\sun}$ at 0$\arcsec$.2, which is about three orders of magnitude below the accretion luminosity inferred for LkCa~15b \citep{Sallum2015}.

Table 1 lists the 40 sources from the SEEDS study and the studies 
of individual disks that have been observed with angular differential imaging or aperture masking.
Of these 
24 are LMS and 16 are IMS (Column 2). 
Of the 40 sources, 20 have been classified as transition disks (Column 4).
The 4 sources with two-arm spirals---the other signature of high-mass giant planets---are all transition disk sources. Column 5 provides the cavity size of the transition disks. Columns 6 and 7 list the inner and outer extent of the region of the disk that has been imaged. Columns 8-10 present the NIR magnitude of the stars, columns 11-13 present the contrast limits achieved at 0$\arcsec$.25, and columns 14-16 present the 
measured value or upper
limit on $\rm \lambda L_{\lambda}$ of a companion in each band. Assuming the colors of an accreting disk source (see Section 3), these 
upper limits are sufficient to detect a companion as bright as 0.1$\rm L_{\sun}$ and sometimes much fainter.

The above studies have tended to focus on disks 
with spiral arms, gaps, and cavities, which are potential 
signposts of planets \citep[e.g.][]{Dong2015a, Dong2015b}. 
Unlike direct imaging searches for planets around stars 
without a disk, searches for planets in disk-bearing 
systems are complicated by the effect of the protoplanetary 
disk material on the detectability of embedded planets  \citep[e.g.,][]{maire2017}. 
As shown in the Table, most such 
efforts have yielded non-detections. 
For example, \citet{maire2017}
and \citet{canovas2017} reported the non-detection of planets in the 
SAO 206462 and 2MASS J1604 disks, respectively, and placed 
an upper limit of a few Jupiter masses on the mass of putative planets  
at $r\gtrsim$ 100 AU in the former and at $r\gtrsim30$ AU 
in the latter. 

Among these 40 systems, the most secure detection of accreting protoplanets is in PDS~70 \citep{Keppler2018, haffert2019}. 
PDS~70 has detected companions at $20.6\pm1.2$\,au and $34.5\pm2.0$\,au \citep{haffert2019}.   
The companions have been imaged in the NIR 
\citep{Keppler2018}, H$\alpha$ \citep{Wagner2018, haffert2019}, and their associated circumplanetary disks may have been discovered in mm continuum emission \citep{Isella2019}. From their H$\alpha$
measurement, \citet{Wagner2018} estimate a planetary accretion rate 
of $10^{-11\pm1}~M_{\odot}~yr^{-1}$. PDS~70 is a weak lined T Tauri star 
whose H$\alpha$ equivalent width is $\rm \sim 2\AA$ \citep{Greg2002}. 
Adopting the stellar parameters presented in \citet{Long2018} and the prescription for converting line luminosity
to accretion luminosity given in \citet{Fang2009}, we arrive 
at a stellar accretion rate of $8\times10^{-11}~M_{\odot}~yr^{-1}$. For such low
rates of accretion, it is possible that most of the H$\alpha$ emission 
arises from chromospheric activity, so this accretion rate should be taken 
as an upper limit. Over 5\,Myr, only 0.4\,M$\rm_J$ would be accreted
by the star at this rate. 
If they are hot start planets,
the masses of PDS~70b and PDS~70c are estimated to be 4 -- 17 M$\rm_J$ and 4 -- 12 M$\rm_J$ respectively \citep{haffert2019}
and the accretion rate would 
have been substantially higher in the past.
There are significant 
uncertainties in both the planetary and stellar accretion rates, yet the 
derived values are consistent with a reasonable fraction 
of the accreting material through the disk being captured by the planet. 
 
Multiple orbiting planetary companions have been reported in 
association with LkCa~15 (\citealt{Kraus2012, Sallum2015}),
its three planetary candidates detected at $\sim 15$--20\,au 
in the inner cavity of its disk 
using NIR sparse aperture masking (SAM) \citep{Sallum2015}. 
One of these sources has been imaged in H$\alpha$, from which the accretion rate onto the planet has been estimated. The
infrared colors and H$\alpha$ emission are consistent with a planet mass 
times accretion rate $M_p\dot{M_p}\sim3-10\times10^{-6}~M^2_J~yr^{-1}.$ 
For a Jovian mass companion, this translates to an accretion rate of 
$\sim3-10\times10^{-9}~M_{\odot}~yr^{-1}$, which is within a factor of 
a few of the stellar accretion rate measured 
for LkCa~15 ($3.6\times10^{-9}~M_{\odot}~yr^{-1}$; \citealt{Ingleby2013}). A subsequent study found that 
the H$\alpha$ luminosity of the companion appears to vary, 
indicating variable accretion \citep{Mend2018}. 
While there is significant uncertainty about the accretion rate onto 
the planet, the derived values are largely consistent with the expectation
that a non-negligible fraction of the mass accreting through the disk makes 
it across the planet's orbit into the inner disk \citep{Lubow2006}.
While the detection of orbital motion  in data taken over 6 years supports the interpretation that the emission arises from massive planetary companions (\citealt[]{Sallum2015}, Sallum et al. in prep.),
other studies of LkCa~15 that use direct imaging techniques rather 
than SAM find structures that are more consistent with emission 
from an inner disk \citep[]{currie2019, Thalmann2016}. 
Further study is needed to 
understand the properties of the orbiting emission sources.

One or more circumplanetary disks may have been detected in the 
HD~100546 system \citep{quanz2013, brittain2013, brittain2014, brittain2015, Brittain2019, Currie2015, currie2017, Lisk2012}.
This system hosts a 
source of $5\micron$ CO fundamental emission located 
$\sim$12~au from the star, close to the inner rim of the outer disk,
whose orbit has been followed for 
15~yrs. The CO flux is consistent with emission from 
a circumplanetary disk with a radius of $\sim$0.3~AU if we assume 
the emitting gas is optically thick and at the same temperature as 
the circumstellar gas near the disk edge (1400~K; \citealt{brittain2013}). 
The temperature and emitting area are 
consistent with the theoretically 
predicted thermal properties of circumplanetary disks surrounding 
giant planets. \citet{szulagyi2014-acc} reported a disk temperature of
$\sim$2000\,K out to a Hill radius R$\rm_{Hill}\sim$0.8~au for their case of a 
10~M$\rm_J$ planet at 5~au (see also \citealt{szulagyi2017-temp, szulagyi2017-thermo}). 
Similar temperatures have been reported for 
the inner circumplanetary disk in other three-dimensional radiation 
hydrodynamical simulations \citep{Klahr2006, Gressel2013}. 

Confirmation of this 
source by direct imaging remains ambiguous. \citet{Currie2015} report the possible 
presence of a point source at the expected location of the CO emission, although 
the point source was
not confirmed in subsequent observations \citep{Follette2017, Rameau2017}. 
One reason for the differing results may be because the point source 
fell behind the coronagraphic mask
in the later observations \citep{currie2017}. Attempts to detect the source with 
ALMA continuum imaging have also failed \citep{Pineda2019}, although it is likely that the pressure bump at the inner edge of the outer disk strongly filters out 
millimeter-sized dust grains 
and prevents them from reaching the circumplanetary disk. Moreover, grains that make it to the circumplanetary disk are expected to drift inward quickly \citep{zhu2018}, enhancing the gas-to-dust ratio 
of the circumplanetary disk and leading to weak millimeter continuum emission. 
Thus the status of this 
possible planet remains uncertain. In addition
to the source at $\sim$12~au, an extended source of IR emission has been 
reported at $\sim$50~au \citep{quanz2015ApJ}, and a third
millimeter continuum point source 
has been detected at 5.6~au with ALMA \citep{Perez2019}. \footnote{A
companion candidate has very recently been detected in LBT $L'$ and $M$-band imaging of the MWC 758 disk \citep{Wagner2019}. Assuming the emission is entirely photospheric, the planet is estimated to be 2-5 $M_{\rm J}$ in mass in a hot-start scenario. No accretion signature has been detected from the candidate so far. As confirmation of the candidate is currently underway (K. Wagner, priv. comm.), we do not include this candidate in our study.}

In summary, 
among the 20 sources studied to date that show signposts of giant planets 
(large cavities or two-arm spirals), 
accreting gas giant planets have been detected in
at least 1 young LMS system (PDS~70) and possibly 1 additional young LMS system (LkCa~15) and 1 young IMS system (HD~100546; Table 2). 
These results correspond to a detection rate 
of 5--15\% among this select group, 
whereas we would have expected to detect one or more giant planet companions in every system if all transition disks host multiple high-mass giant planets that radiate away their accretion energy as they form.
In addition, 
no planetary companions have been reported in association with the 
remaining 20 non-transition disk sources in Table 1. 

These results indicate that the incidence rate of bright, detectable giant planet companions among all stars is very low. 
 That is, among young LMS, only 5-15\% of their protoplanetary disks have 
transition disk SEDs 
(e.g., \citealt{muzerolle2010, Furlan2011}), 
and of the 12 young LMS transition disks in Table 1, 
only 1, possibly 2, have a bright detected companion, for a detection rate of 8-16\%. The product of these 
two rates implies an incidence rate of accreting giant planetary companions to young LMS of $\sim 1$\%. 
Similarly, 
among IMS protoplanetary disks, 
$\sim 40$\% have transition disk SEDs (see Brittain et al.\ 2019, in preparation) 
and $\sim 20$\% have two-arm spirals
\citep{dong2018a}, whereas none or possibly one (HD~100546)
of the 8 young IMS transition disks in Table 1 
 have a bright detected companion, for a detection rate of 0--12\%. The product of these 
two rates implies an incidence rate of accreting giant planets of 0--5\% for young IMS. 
In other words, the incidence rate of currently detectable
accreting giant planet companions 
among all protoplanetary disks is likely of the order of a few percent.

\section{Toy Model of Episodic Accretion}

It is perhaps surprising that the planet detection rate at these
very young ages is so low, especially if most forming planets
radiate away their accretion energy as they form 
(i.e., they are cold-start planets) and are expected to be bright in their mass-building phase.
Here we explore the role episodic accretion may play in 
accounting for the dearth of 
bright young systems observed among these disks. 

\citet{Lubow2012} have noted that circumplanetary disks, 
like the more extended circumstellar disks in which they reside,  
are likely to harbor significant dead zones, i.e., regions that 
are insufficiently ionized to participate in accretion 
via the magnetorotational instability (MRI). 
As a result, as it is fed material from the circumstellar disk, 
the circumplanetary disk will grow in mass until it becomes 
gravitationally unstable. 
By driving turbulent heating (and thus ionization), 
gravitational instability warms the disk until it is sufficiently 
ionized thermally to drive accretion via the MRI, a process  
they refer to as the gravo-magneto instability \citep{Armitage2001, Zhu2009, martin2011}. 
Their model predicts outburst rates ranging from $0.04\rm M_J~yr^{-1}$ 
-- $2 \rm M_J~yr^{-1}$, and quiescent rates ranging from $4\times10^{-7} \rm M_J~yr^{-1}$ 
-- $3\times10^{-5} \rm M_J~yr^{-1}$, 
with accretion outbursts that last for several years
occuring every $10^4 - 10^5$ yrs 
\citep{Lubow2006}.

To compare the observed properties of detected accreting 
giant planets with the predictions of episodically accreting 
circumplanetary disks, 
we extend the work of \citet{Lubow2012} by considering 
the effect of non-steady state disk accretion and the growing
mass of the planet. To set the rate at which the circumplanetary 
disk is fed, 
we first consider a circumstellar disk with an accretion rate 
that declines with time as t$^{-3/2},$ mimicking the decline 
in the average measured stellar accretion rate over Myr 
timescales\footnote{As noted by \citet{Hartmann2016}, although measured stellar 
accretion rates follow this declining trend, 
at stellar ages beyond $\sim 3$\,Myr most young stars have        
undetectable accretion. Our assumed rate of 
decline therefore overestimates the stellar accretion rate at late 
times, although this flaw has little impact on our 
results qualitatively. 
The effect of a steeper decline in the accretion rate is to fill the circumplanetary disk more slowly at later times resulting in more time between outbursts.} (see for example \citealt{Sicilia2005}), such that 
\begin{equation}
    \dot{M}(t)=\dot{M}(t_0)(t/t_0)^{-3/2}. %- 10^{-9} M_{\sun}\ \rm yr^{-1},
\end{equation}
\noindent We adopt $t_0=1$~Myr and an initial accretion rate of 
$\dot{M}(t_0)=10^{-8}~M_{\odot}~yr^{-1},$ 
an accretion rate typical of young T Tauri stars 
(e.g., \citealt{Hartmann1998}). 
We then embed in the circumstellar accretion disk a forming 
giant planet with a mass $M_p=20M_{\earth},$ roughly the mass 
at which runaway accretion begins \citep[e.g.,][]{DAngelo2010}. 

From a theoretical perspective, 
the fraction of the accreting circumstellar disk material that 
is captured by the planet, $f(M_p)$, depends on a variety of 
parameters including $q$, the mass ratio of the planet and star, 
the orbital eccentricity of the companion, and the viscosity 
of the gas. Estimates from 2-dimensional simulations indicate 
that a forming gas giant planet captures 75\%--90\% 
of the accreting material \citep{Lubow2006}. However, the value 
is uncertain because the flow onto the planet and circumplanetary 
disk is intrinsically 3-dimensional 
\citep{Ayliffe2012-env, Batygin2018, Tanigawa2012, Fung2015, szulagyi2016-disc}.

Observationally, 
there are few constraints on the captured fraction, 
although we might attempt to infer its typical value 
from the ratio of the stellar and planetary accretion rates 
for the few accreting companions detected to date. 
As described in \S2, 
the properties of the accreting companions to 
LkCa~15 and PDS~70 indicate that a significant portion 
of the material accreting through the disk is captured by the 
companion ($\sim$50\% and $\sim$10\% respectively), while 
only a small fraction of the mass is captured 
in the case of the companion to HD~142527 ($\sim$10$^{-3}$). 
These values are uncertain because the scaling relationship 
between H$\alpha$ emission and planetary accretion rate 
may be in error, the accretion onto the planet may not be 
in steady state \citep[e.g.,][]{Mend2018}, and/or 
the H$\alpha$ line may be more heavily extincted by 
circumplanetary matter than estimated. 

Given the uncertainties surrounding the captured fraction, 
we parameterize it as 
\begin{equation}
    f(M_p)=\frac{1}{1+a/q},
\end{equation}
\noindent where $a$ is a free parameter, 
and $q=M_p/M_\star$ is 
the ratio of the companion mass to the mass of the central star. We arrived at this
functional form and the value $a = 10^{-4.5},$ by fitting the relationship between $f$ and $q$ found in 
the 2-d hydrodynamical simulations presented by \citet{Lubow2006}. 
Here we explore how the efficiency factor affects the duty cycle of 
accretion outbursts. 

As material is captured by the forming planet, it fills 
a circumplanetary disk, which accretes onto the planet 
at a rate that depends on $\zeta,$ the mass ratio 
of the circumplanetary disk to the forming planet. 
Here we assume 
\begin{equation}
 \begin{split}
    \dot{M}_{p} &= 5\times10^{-4} M_{\rm J}~{\rm yr}^{-1}; {\rm for\ } \zeta \gtrsim 0.1 \\
                     &= 5\times10^{-8} M_{\rm J}~{\rm yr}^{-1}; {\rm for\ } \zeta \lesssim 0.1. \\
 \end{split}
\end{equation}
\noindent Although these rates are highly uncertain as well, 
they are based on measured accretion rates for young stars. While \citet{Lubow2012} adopt much higher 
accretion rates in their outburst, \citet{hall2019} find that in their models that for marginally gravitationally unstable disks, the accretion rate is of order 10$^{-7}$ M$_{\sun}$ yr$^{-1}$ which we adopt for our model. 

In quiescence, we assume that the accretion rate onto the planet is reduced by four orders of 
magnitude, consistent with the models of  
\citet{Lubow2012}. We allow the 
model to run until $q = 5\times10^{-3},$
at which point the forming companion halts accretion 
into its circumplanetary disk \citep{lubow1999,Lubow2006}. 

With these assumptions, the planet grows to 9~$M_J$ in about 3~Myr 
(Fig.\ 1b). 
The accretion luminosity of the planet alternates between 
$\sim 4\times 10^{-5}$\,L$_{\sun}$ and $\sim$0.4\,L$_{\sun}$. Outbursts occur every 1500~yr and last $\sim$30~yr; thus the circumplanetary 
disk spends $\sim$2\% of the runaway accretion phase undergoing an 
outburst (Fig.\ 1a). 
Even in quiescence, the accretion luminosity exceeds the luminosity 
of a cold start planet ($\sim 10^{-6}L_{\sun}$; Fortney et al.\ 2008).
We can compare the accretion luminosity to the integrated luminosity of the detected companions orbiting 
HD~142527, LkCa~15, and PDS~70. We fit the disk accretion models by \citet{Zhu2015} to the measured photometry for each source (Figure 2) and 
find that their luminosities range from $10^{-4}$L$_{\sun}$ to 0.6\,L$_{\sun}$ (see Table 3 
for the properties of the companions). The gray bars in figure 2 show the range of upper limits on $\rm \lambda L_{\lambda}$ of the sample in Table 1. Adopting 
the colors from the \citet{Zhu2015} model, we find that the published observations are sufficiently sensitive to detect accreting circumplanetary disks with an integrated 
luminosity of $\rm 10^{-4} - 0.1~L_{\sun}$.

Thus, studies to date have imaged 20 disks with transition disk SEDs 
(4 of which have two-arm spirals) with sufficient sensitivity to 
detect a companion as bright as 0.1\,L$_{\sun}$ (see \S 2).  
%As a result, 
If each of the 20 disks harbors one accreting gas giant planet,
and their circumplanetary disks are in outburst 2\% of the time, 
we have a 33\% chance of catching one or more circumplanetary disks in 
outburst. 
\citet{dodson2011} argued that the large cavity 
sizes of transition disks required the presence of 
multiple multi-Jupiter mass planets within the optically 
thin region. 
If each transition disk harbors two (three) accreting planets 
(and the observations probe the orbital radii of both/all planets),
we have an 55\% (70\%) chance of catching one or more planets in outburst. 

Interestingly, one companion as bright as 
our outburst luminosity ($\sim 0.6\,L_\sun$) has been 
detected to date, HD~142527B. 
In \S 4 we explore the possibility 
that HD~142527B is a planet surrounded 
by a circumplanetary disk in outburst. 
More generally though, the toy model suggests that episodic 
accretion can plausibly explain the low detection rate of 
accreting planetary companions to young stars. 
Future observational constraints on the incidence rate of giant 
planets and their luminosity in the accretion phase can 
directly constrain the duty cycle of episodic accretion 
and the outburst accretion rate. 

In the episodic accretion picture, forming planets 
would present two faces to the world. Most of the time, 
they are accreting very slowly, they will look like 
a faint ``cold start'' planet surrounded by a quiescent disk. 
A small fraction of the time (a few \%), when in outburst, 
the emission from system will be dominated by emission 
from the disk, as in an FU~Ori object, with the color 
and luminosity of an M star. In the next section, 
we discuss whether the directly imaged companion to HD~142527 
could be in the latter state.

\section{HD~142527: Stellar or Planetary Companion? }
The directly imaged companion to 
the young F6 star HD~142527 (2.0$\pm$0.3~M$_{\sun}$) 
has been detected through its NIR continuum and H$\alpha$ emission 
at a separation of $\sim 13$\,au from the star 
\citep{Biller2012, Close2014}. 
A spectrum of the companion in the $H$- and $K$-bands has 
also been acquired \citep{Christiaens2018}. 
These observations have been interpreted as emission 
from a young M2.5 star 
with a mass of $\sim$0.1--0.4\,M$_{\sun}$ that is accreting at a rate 
$6\times10^{-10}~M_{\odot}~yr^{-1}$ 
\citep{Close2014, Lacour2016, Biller2012, Christiaens2018}, 
a tiny fraction of the accretion rate onto this 5.0$\pm$1.5~Myr old star 
($\sim 2\times 10^{-7}$~M$_{\sun}$~yr$^{-1}$) \citep{Mend2014}. 

Studies that have used dynamical arguments to constrain 
the mass of the companion favor a stellar companion, although 
the constraints admit the possibility of one or more planetary 
mass companions instead 
(\citealt{Price2018}; \citealt{Claudi2019}; see \S 5 for further details). 
In the absence of strong evidence to the contrary,  
we here consider the possibility that the companion is 
a planetary mass companion surrounded by a 
circumplanetary disk undergoing an accretion outburst. 
Earlier studies have suggested that 
the NIR spectrum of an active accretion disk can mimic that of 
an M-star (e.g., 
\citealt{Herbig1977}; see also \citealt{Zhu2015}).

Figure 2a compares the IR photometry of the companion  
\citep[black points; ][]{Lacour2016} 
with the model spectrum of an accreting circumplanetary disk 
(black spectrum). 
In the model, the circumplanetary 
disk accretes at a constant rate onto the planet. 
The effective temperature of the circumplanetary disk 
is the standard steady optically thick accretion disk temperature.  
The vertical temperature dependence of the disk atmosphere 
at each radius is calculated using the gray-atmosphere 
approximation in the Eddington limit, adopting the Rosseland mean optical depth.  
With the temperature determined at each radius, the SED of the local annulus is 
calculated. By summing the SEDs from different annuli, the SED of the accretion 
disk is obtained \citep{Zhu2015}. 

The other model parameters---in addition to the disk 
accretion rate and the range of disk radii that contribute to 
the emission---are the disk inclination and the reddening to the disk. 
The inclination of the outer disk around HD~142527 is 
$\sim$20-30$\degr$ \citep{Pont2011,Casassus2013}, whereas the 
inner circumstellar disk is inferred to have an inclination 
$\sim 70\degr$ relative to the outer disk \citep{Marino2015}. 
If the companion has a disk, it is not clear what 
the orientation of the disk would be, so its inclination is 
treated as a free parameter in our fit. 
The planet and circumplanetary disk are expected to be 
embedded in a circumplanetary envelope \citep{Tanigawa2012, Gressel2013, szulagyi2016-disc} 
that may extinct and redden the emission from the planet+disk 
by an unknown amount. 
%, the local reddening 
%of the disk from a circumplanetary envelope is unclear.  
%Thus we deredden the photometry with 
%
We therefore assume a parameterized reddening law 
$A_{\lambda}=A_J( \lambda/1.235)^{-\alpha}$ 
where $A_J$ and $\alpha$ are free parameters. 

In the model calculation, 
the inclination correction is treated simply as a multiplicative 
factor that corrects for the projected disk emitting area. 
%
%(which has the effect of constant flux offset) and reddening 
%and compared to the model.  
%
We obtain a reasonable fit with 
a disk that extends from an inner radius of $R_{in}=1.5R_J$ 
to an outer radius of $R_{out}=15R_J$, an accretion rate 
$\dot{M_p}$ such that 
$M_p\dot{M_p}=10^{-2}~M_J^2~yr^{-1}$, where 
$M_p$ is the planetary mass, 
%The data are well fit with model 
and 
extinction and inclination parameters of 
$A_J=1.6$, $\alpha=1.7$, and $i=70\degr$ (Fig.\ 2a).

To compare the model to the NIR spectrum of the companion  
\citep{Christiaens2018}, we assume the 
absolute flux calibration of the photometry is more reliable than the spectrum, so 
we scale the spectrum to be consistent with the photometry.  
The model and scaled spectrum agree well (Figure 2), 
although there is a slight discrepancy in the $K$-band region. 
Such differences are not surprising, because of the simple 
assumptions made in the model. For example, the model 
assumes a simple vertical
temperature structure (based on a grey atmosphere with no 
temperature correction; \citealt{Zhu2015}). Any deviation from the standard 
radial disk temperature profile assumed here will also alter 
the SED. 
Furthermore, the opacity of the model is fixed at solar abundance 
and an adopted mean opacity, so any changes 
in the elemental composition of the accreting material 
would affect the SED. For example, any trapping of large grains in 
pressure bumps in the circumstellar disk will reduce the dust content of the material accreting into circumplanetary disk. 

These results show that the emission properties of the companion to HD~142527 
are plausibly those of a giant planet surrounded by a 
circumplanetary disk that is undergoing an accretion outburst  
and whose emission dominates the emission from the planet. 
If the companion is in fact a 10~$M_J$ planet surrounded by a circumplanetary disk, then 
our fit value of 
$M_p\dot{M_p}=10^{-2}~M_J^2~yr^{-1}$ 
implies a planetary accretion rate of 
$10^{-6}~M_{\odot}~yr^{-1},$ roughly an order of magnitude
higher than the stellar accretion rate for this system \citep{Mend2014}.

\section{Discussion}
In \S 1, we argued that the high incidence rate 
of disk substructure indicating the presence 
of massive ($\sim 5 M_J$) giant planets (e.g., two-arm spirals 
and transition disk morphologies associated with $\sim$ 10--20\% 
of disk-bearing stars) 
coupled with the lack of directly imaged ``hot start'' 
planets in this mass range at $> 10$\,Myr ages, 
suggests that giant planets form ``cold start'', 
i.e., 
that planets radiate away much their accretion energy 
in the accretion phase.
The lack of evidence for energy loss in this form, i.e., 
the ready detection of bright, accreting planets in the 
pre-main-sequence phase ($\sim 1$\, Myr), led us to 
propose that planets accrete their mass episodically, 
through punctuated outbursts of accretion in the 
runaway gas accretion phase. 

This interpretation may appear too glib when we consider 
that some of the first directly imaged planets, those orbiting 
$\beta$~Pic and HR~8799, appear to be ``hot start'' 
planets. Dynamical constraints on their masses, 
coupled with their observed luminosities, are consistent 
with the predictions of hot start models \citep[][]{Fab2010,wang2018}. While these 
planets were first reported many years ago \citep{Marois2008,Lagrange2009},
subsequent discoveries of directly imaged planets have been 
few and far between \citep{Bowler2016,Stone2018, Nielsen2019}.

One interpretation of the discrepancy between these results 
is that planet formation proceeds through 
multiple pathways. Planet formation via gravitational 
instability, which favors high mass planets ($> 10 M_J$), 
is expected to produce hot start planets, while 
core accretion, which favors low mass planets ($< 10 M_J$), 
is expected to produce 
colder start planets. Both pathways may lead to 5--15 $M_J$ 
planets at the orbital separations probed by direct imaging, 
with the brighter hot start planets readily detected 
($\beta$~Pic, HR~8799) 
and the (possibly more numerous) cold start planets as yet unprobed. 

The episodic accretion scenario described in \S3
predicts that accreting 
planets will come in two flavors: (1) outbursting 
systems with such high disk accretion rates that their emission is dominated by the circumplanetary disk and 
has the color of low mass stars 
and (2) systems that are faint because 
the circumplanetary disk is inactive. 
At a planet mass of $\sim 3 M_{\rm J}$---the mass 
of the multiple planets invoked by \citet{dodson2011} 
to explain the large cavities 
of transition disks---the outbursting state in our toy model corresponds to $M_p\dot{M_p}$=$\rm 1.5\times10^{-3}~M^2_J~yr^{-1}$ and is sufficiently bright to have 
been detected by the 40 published observations of disks (Figure 2b, Table 1).
The quiescent state for such a $3 M_{\rm J}$ planet corresponds to 
$M_p\dot{M_p}$=$\rm 1.5\times 10^{-7}~M^2_J~yr^{-1}$ or about two orders 
of magnitude fainter 
at $K$-band 
than the most sensitive published observations to date (Fig.\  2). 
As with young stars, there is likely a large range of accretion 
rates that represent the quiescent, steady-state rate of circumplanetary disks. 
Because quiescent disks are faint in the NIR and 
emit more of their energy in the MIR, searches at longer wavelengths 
may be better able to detect quiescent disks
%The photometry compiled is in the NIR ($J$, $K_s$, and $L$ bands), 
%however, at lower accretion rates, the brightness of the models peak in the MIR 
(Fig.\ 2; see also \citealt{Szulagyi2019}).

While the emission properties of HD~142527b are plausibly consistent with those of a
circumplanetary disk in outburst (Fig.\ 2b), the companions PDS~70bc and 
LkCa~15bc have fluxes between the outburst and quiescent states of our model.
We hypothesize that these values 
reflect the upper limit of the range of quiescent states rather than the 
lower range for outbursting disks, because it is unlikely that two 
circumplanetary disks in a given system would undergo an outburst simultaneously.

Other authors have described how the HD~142527 system 
properties are consistent with a stellar, rather than 
planetary, companion \citep[e.g.,][]{Biller2012, Lacour2016, Christiaens2018}. 
A stellar companion is consistent
with many detailed aspects of the system: 
the multiple spiral arms in the outer disk at $r\gtrsim100$~au 
\citep{Fukagawa2006, Canovas2013}, 
the large central clearing in the disk \citep[$\sim$1\arcsec radius;][] {Fukagawa2013}, 
the azimuthal asymmetry in the outer disk \citep{Casassus2013}, 
and evidence for a (spatially unresolved) inner disk that 
is highly misaligned with the outer disk \citep{Marino2015}. 
\citet{Price2018} proposed that these properties can be 
explained by a single companion that has a mass of 
0.4 M$_{\sun}$ and an unusual orbit, 
with  both high eccentricity ($e$=0.6--0.7) and 
an inclination that is almost polar with respect to the outer disk. 

\citet{Price2018} state that their results are not strongly sensitive 
to the companion mass in the range they studied, and 
a lower, planetary mass companion 
(e.g., $\sim 10\,M_J$) is not clearly excluded. 
\citet{Lacour2016} and 
\citet{Claudi2019} have shown that both the mass and the orbit of 
the companion are highly uncertain.
For example, \citet{Claudi2019}
placed a dynamical constraint on the mass of the companion of $0.26^{+0.16}_{-0.14}$~M$_{\sun}$. 
Thus the analysis favors a stellar companion but does not 
rule out a $10 M_J$ companion. 
It is also unclear whether all of the circumstellar disk properties 
are the result of a single companion. A disk with a 
very large inner hole, like that of HD~142527, could signal the presence of 
multiple giant planet companions \citep{zhu2011,dodson2011}; 
perhaps the best example 
of this is PDS~70bc \citep{haffert2019}.
Thus available data allow for the possibility of a 
planetary mass companion.

The episodic accretion picture 
can be tested by searching for orbiting companions
in their more typical, quiescent state, with deeper 
high contrast imaging than has been performed to date. Orbiting companions may also be identified through
submillimeter continuum imaging of circumplanetary disks, 
infrared imaging and spectroscopy of circumplanetary disks, 
and Gaia astrometry. 

ALMA may be able to detect quiescently accreting circumplanetary disks 
through their dust emission \citep[e.g.,][]{zhu2018, Boehler2017, Isella2014}, however
the strength of the emission depends on the extent to which 
disk solids are filtered out by gap edge of the outer disk  \citep[e.g.,][]{rice2006}
before reaching the 
circumplanetary disk. Efficient filtering or rapid growth 
into large solids will reduce the circumplanetary disk optical depth, 
potentially compromising the detectability of the dust emission signature. 
Perhaps as a result, circumplanetary disk detections 
with this approach have been rare to date despite multiple attempts. 
An exciting possible detection has been reported by \citet{Perez2019} in the HD~100546 system. 

Circumplanetary disks can also be detected and studied 
in the infrared. Quiescent circumplanetary disks can potentially be 
detected through gas emission features from their atmosphere, 
e.g., with spectroastrometry of 5 $\micron$ CO fundamental emission as in HD~100546 \citep{Brittain2019}.
Future observations can potentially also distinguish between the 
stellar vs.\ outbursting circumplanetary disk explanations for the detected 
companion to HD~142527, e.g., by  
using infrared spectroscopy to measure 
the gravity of the companion. 
Compared to the dwarf-like gravity expected for a young low-mass companion, 
accretion disks in outburst are expected to show giant-like 
gravity, e.g., in their 2.3\,\micron\ CO overtone absorption 
(e.g., FU Ori; \citealt{Kenyon1995}). 
More sensitive searches for companions to much larger samples of 
young stars will be possible with imagers on 30-m class
telescopes. \citet{Zhu2015} and \citet{Szulagyi2019} suggest that the mid-infrared observations
will be more sensitive to the presence of forming planets 
and their circumplanetary disks than the near-infrared
observations pursued thus far.
Mid-infared observations are not only better matched 
to the spectral region where much of the flux is emitted, 
they are also less sensitive to potential obscuration by dust in the 
surrounding protoplanetary disk environment. 

One might wonder whether significant obscuration can explain 
the low detection rate of forming planets, 
obviating the need for episodic accretion. 
At first glance, it seems unlikely that extinction by the circumstellar disk 
can account for the observational results. 
Simulations of gap opening show that even relatively low-mass planets 
($\sim$0.5~M$\rm_J$) in a modest viscosity environment ($\alpha=10^{-3}$)
are able to clear a gap to $\sim$3\% of its original density \citep{Fung2014}. 
For a gap in a transition disk reduced in column density by a roughly a factor of 100-10,000 
relative to the MMSN \citep{vanderMarel2016, Furlan2011},
this results in negligible near-infrared (i.e., K-band) extinction of the forming planet ($\lesssim$~0.05~mag).  
Once the planet grows to greater than a Jupiter mass, this problem becomes even less severe, so it is plausible 
that circumstellar disk extinction does not account for the dearth of forming planets imaged in disks. Longer wavelength observations can test this picture.

One particularly promising way to identify 
weakly or non-accreting gas giant
planets in disks is through the use of {\it Gaia} astrometry. This method
has already been used to measure the mass of the gas giant 
orbiting the A-star $\beta$~Pic. 
The current data from {\it Gaia} and {\it Hipparcos} are able to detect 
the stellar reflex motion induced by an orbiting 
11--13 M$\rm_J$ planet in an A star 20\,pc away with a precision of 3~M$\rm_J$ \citep{Snellen2018, Dupuy2019}  
A similar approach can be used to measure the mass of the 
companion to HD~142527 and test our scenario that it is an 
accreting planet rather than a star.
Once the full {\it Gaia} time baseline becomes available, 
along with the expected 100-fold  increase in  
astrometric precision, it should be possible 
to detect the stellar reflex motion of a sub-Jovian mass planet ($\rm \geq0.3~M_J$) at the 
distance of HD~142527 (157\,pc) or definitively rule out a substellar mass for the
companion. 

Finally, an observational campaign focused on a larger sample 
of sources, especially those with dynamical signatures of 
massive planets (large and deep cavities or two-arm spirals), will  constrain the duty cycle and magnitude of outbursts. 
The fraction of bright circumplanetary disks detected in 
outburst reflects the duty cycle of the outbursts, which 
depends on the rate at which the circumplanetary disk grows 
and the rate at which it empties. As the rate of accretion onto the 
circumplanetary disk increases, so does the frequency of outbursts, 
if their magnitude and duration remain the same. 
If outbursts are more violent than we assume (with planetary accretion rates of $\dot{M}_{p}\gtrsim10^{-6}~M_{\odot}~yr^{-1}$), they will
be brighter and the duty cycle lower.
Surveys that report upper limits on the luminosity of companions 
in systems that are expected to harbor gas giant planets are crucial for 
advancing our understanding of this very important phase of 
gas giant planet growth.

\acknowledgments

We are grateful to Scott Kenyon for comments on an early version of this manuscript. 
This work was performed in part at the Aspen Center for Physics which is 
supported by the National Science Foundation grant PHY-1607611. Work by 
SDB was performed in part at the National Optical Astronomy Observatory. 
NOAO is operated by the Association of Universities for Research in Astronomy 
(AURA), Inc. under a cooperative agreement with the National Science Foundation. 
SDB also acknowledges support from this work by NASA Agreement No. NXX15AD94G; 
NASA Agreement No. NNX16AJ81G; and NSF-AST 1517014.

%\nocite{*}
\bibliography{cite}

\begin{thebibliography}{}
\expandafter\ifx\csname natexlab\endcsname\relax\def\natexlab#1{#1}\fi

\bibitem[{{Akiyama} {et~al.}(2016){Akiyama}, {Hashimoto}, {Liu}, {Li},
  {Bonnefoy}, {Dong}, {Hasegawa}, {Henning}, {Sitko}, {Janson}, {Feldt},
  {Wisniewski}, {Kudo}, {Kusakabe}, {Tsukagoshi}, {Momose}, {Muto}, {Taki},
  {Kuzuhara}, {Satoshi}, {Takami}, {Ohashi}, {Grady}, {Kwon}, {Thalmann},
  {Abe}, {Brandner}, {Brand t}, {Carson}, {Egner}, {Goto}, {Guyon}, {Hayano},
  {Hayashi}, {Hayashi}, {Hodapp}, {Ishii}, {Iye}, {Knapp}, {Kand ori},
  {Matsuo}, {Mcelwain}, {Miyama}, {Morino}, {Moro-Martin}, {Nishimura}, {Pyo},
  {Serabyn}, {Suenaga}, {Suto}, {Suzuki}, {Takahashi}, {Takato}, {Terada},
  {Tomono}, {Turner}, {Watanabe}, {Yamada}, {Takami}, {Usuda}, \&
  {Tamura}}]{Akiyama2016}
{Akiyama}, E., {Hashimoto}, J., {Liu}, H.~B., {et~al.} 2016, \aj, 152, 222

\bibitem[{{ALMA Partnership} {et~al.}(2015){ALMA Partnership}, {Brogan},
  {P{\'e}rez}, {Hunter}, {Dent}, {Hales}, {Hills}, {Corder}, {Fomalont},
  {Vlahakis}, {Asaki}, {Barkats}, {Hirota}, {Hodge}, {Impellizzeri}, {Kneissl},
  {Liuzzo}, {Lucas}, {Marcelino}, {Matsushita}, {Nakanishi}, {Phillips},
  {Richards}, {Toledo}, {Aladro}, {Broguiere}, {Cortes}, {Cortes}, {Espada},
  {Galarza}, {Garcia-Appadoo}, {Guzman-Ramirez}, {Humphreys}, {Jung}, {Kameno},
  {Laing}, {Leon}, {Marconi}, {Mignano}, {Nikolic}, {Nyman}, {Radiszcz},
  {Remijan}, {Rod{\'o}n}, {Sawada}, {Takahashi}, {Tilanus}, {Vila Vilaro},
  {Watson}, {Wiklind}, {Akiyama}, {Chapillon}, {de Gregorio-Monsalvo}, {Di
  Francesco}, {Gueth}, {Kawamura}, {Lee}, {Nguyen Luong}, {Mangum}, {Pietu},
  {Sanhueza}, {Saigo}, {Takakuwa}, {Ubach}, {van Kempen}, {Wootten},
  {Castro-Carrizo}, {Francke}, {Gallardo}, {Garcia}, {Gonzalez}, {Hill},
  {Kaminski}, {Kurono}, {Liu}, {Lopez}, {Morales}, {Plarre}, {Schieven},
  {Testi}, {Videla}, {Villard}, {Andreani}, {Hibbard}, \&
  {Tatematsu}}]{ALMA2015}
{ALMA Partnership}, {Brogan}, C.~L., {P{\'e}rez}, L.~M., {et~al.} 2015, \apjl,
  808, L3

\bibitem[{{Armitage} {et~al.}(2001){Armitage}, {Livio}, \&
  {Pringle}}]{Armitage2001}
{Armitage}, P.~J., {Livio}, M., \& {Pringle}, J.~E. 2001, \mnras, 324, 705

\bibitem[{{Avenhaus} {et~al.}(2018){Avenhaus}, {Quanz}, {Garufi}, {Perez},
  {Casassus}, {Pinte}, {Bertrang}, {Caceres}, {Benisty}, \&
  {Dominik}}]{Avenhaus2018}
{Avenhaus}, H., {Quanz}, S.~P., {Garufi}, A., {et~al.} 2018, \apj, 863, 44

\bibitem[{{Ayliffe} \& {Bate}(2012)}]{Ayliffe2012-env}
{Ayliffe}, B.~A., \& {Bate}, M.~R. 2012, \mnras, 427, 2597

\bibitem[{{Baraffe} {et~al.}(2003){Baraffe}, {Chabrier}, {Barman}, {Allard}, \&
  {Hauschildt}}]{Baraffe2003}
{Baraffe}, I., {Chabrier}, G., {Barman}, T.~S., {Allard}, F., \& {Hauschildt},
  P.~H. 2003, \aap, 402, 701

\bibitem[{{Batygin}(2018)}]{Batygin2018}
{Batygin}, K. 2018, \aj, 155, 178

\bibitem[{{Biller} {et~al.}(2012){Biller}, {Lacour}, {Juh{\'a}sz}, {Benisty},
  {Chauvin}, {Olofsson}, {Pott}, {M{\"u}ller}, {Sicilia-Aguilar}, {Bonnefoy},
  {Tuthill}, {Thebault}, {Henning}, \& {Crida}}]{Biller2012}
{Biller}, B., {Lacour}, S., {Juh{\'a}sz}, A., {et~al.} 2012, \apjl, 753, L38

\bibitem[{{Boehler} {et~al.}(2017){Boehler}, {Weaver}, {Isella}, {Ricci},
  {Grady}, {Carpenter}, \& {Perez}}]{Boehler2017}
{Boehler}, Y., {Weaver}, E., {Isella}, A., {et~al.} 2017, \apj, 840, 60

\bibitem[{{Bowler}(2016)}]{Bowler2016}
{Bowler}, B.~P. 2016, \pasp, 128, 102001

\bibitem[{{Brittain} {et~al.}(2014){Brittain}, {Carr}, {Najita}, {Quanz}, \&
  {Meyer}}]{brittain2014}
{Brittain}, S.~D., {Carr}, J.~S., {Najita}, J.~R., {Quanz}, S.~P., \& {Meyer},
  M.~R. 2014, \apj, 791, 136

\bibitem[{{Brittain} \& {Najita}(2019)}]{Brittain2019b}
{Brittain}, S.~D., \& {Najita}, J.~R. 2019, \apj, submitted

\bibitem[{{Brittain} {et~al.}(2015){Brittain}, {Najita}, \&
  {Carr}}]{brittain2015}
{Brittain}, S.~D., {Najita}, J.~R., \& {Carr}, J.~S. 2015, \apss, 357, 54

\bibitem[{{Brittain} {et~al.}(2019){Brittain}, {Najita}, \&
  {Carr}}]{Brittain2019}
---. 2019, \apj, 883, 37

\bibitem[{{Brittain} {et~al.}(2013){Brittain}, {Najita}, {Carr}, {Liskowsky},
  {Troutman}, \& {Doppmann}}]{brittain2013}
{Brittain}, S.~D., {Najita}, J.~R., {Carr}, J.~S., {et~al.} 2013, \apj, 767,
  159

\bibitem[{{Canovas} {et~al.}(2013){Canovas}, {M{\'e}nard}, {Hales},
  {Jord{\'a}n}, {Schreiber}, {Casassus}, {Gledhill}, \& {Pinte}}]{Canovas2013}
{Canovas}, H., {M{\'e}nard}, F., {Hales}, A., {et~al.} 2013, \aap, 556, A123

\bibitem[{{Canovas} {et~al.}(2017){Canovas}, {Hardy}, {Zurlo}, {Wahhaj},
  {Schreiber}, {Vigan}, {Villaver}, {Olofsson}, {Meeus}, {M{\'e}nard},
  {Caceres}, {Cieza}, \& {Garufi}}]{canovas2017}
{Canovas}, H., {Hardy}, A., {Zurlo}, A., {et~al.} 2017, \aap, 598, A43

\bibitem[{{Canovas} {et~al.}(2018){Canovas}, {Montesinos}, {Schreiber},
  {Cieza}, {Eiroa}, {Meeus}, {de Boer}, {M{\'e}nard}, {Wahhaj},
  {Riviere-Marichalar}, {Olofsson}, {Garufi}, {Rebollido}, {van Holstein},
  {Caceres}, {Hardy}, \& {Villaver}}]{Canovas2018}
{Canovas}, H., {Montesinos}, B., {Schreiber}, M.~R., {et~al.} 2018, \aap, 610,
  A13

\bibitem[{{Casassus} {et~al.}(2013){Casassus}, {Hales}, {de Gregorio}, {Dent},
  {Belloche}, {G{\"u}sten}, {M{\'e}nard}, {Hughes}, {Wilner}, \&
  {Salinas}}]{Casassus2013}
{Casassus}, S., {Hales}, A., {de Gregorio}, I., {et~al.} 2013, \aap, 553, A64

\bibitem[{{Cassan} {et~al.}(2012){Cassan}, {Kubas}, {Beaulieu}, {Dominik},
  {Horne}, {Greenhill}, {Wambsganss}, {Menzies}, {Williams}, {J{\o}rgensen},
  {Udalski}, {Bennett}, {Albrow}, {Batista}, {Brillant}, {Caldwell}, {Cole},
  {Coutures}, {Cook}, {Dieters}, {Dominis Prester}, {Donatowicz}, {Fouqu{\'e}},
  {Hill}, {Kains}, {Kane}, {Marquette}, {Martin}, {Pollard}, {Sahu}, {Vinter},
  {Warren}, {Watson}, {Zub}, {Sumi}, {Szyma{\'n}ski}, {Kubiak}, {Poleski},
  {Soszynski}, {Ulaczyk}, {Pietrzy{\'n}ski}, \& {Wyrzykowski}}]{Cassan2012}
{Cassan}, A., {Kubas}, D., {Beaulieu}, J.-P., {et~al.} 2012, \nat, 481, 167

\bibitem[{{Christiaens} {et~al.}(2018){Christiaens}, {Casassus}, {Absil},
  {Kimeswenger}, {Gonzalez}, {Girard}, {Ram{\'{\i}}rez}, {Wertz}, {Zurlo},
  {Wahhaj}, {Flores}, {Salinas}, {Jord{\'a}n}, \& {Mawet}}]{Christiaens2018}
{Christiaens}, V., {Casassus}, S., {Absil}, O., {et~al.} 2018, \aap, 617, A37

\bibitem[{{Cieza} {et~al.}(2013){Cieza}, {Lacour}, {Schreiber}, {Casassus},
  {Jord{\'a}n}, {Mathews}, {C{\'a}novas}, {M{\'e}nard}, {Kraus}, {P{\'e}rez},
  {Tuthill}, \& {Ireland }}]{Cieza2013}
{Cieza}, L.~A., {Lacour}, S., {Schreiber}, M.~R., {et~al.} 2013, \apj, 762, L12

\bibitem[{{Clarke} {et~al.}(2018){Clarke}, {Tazzari}, {Juhasz}, {Rosotti},
  {Booth}, {Facchini}, {Ilee}, {Johns-Krull}, {Kama}, {Meru}, \&
  {Prato}}]{Clarke2018}
{Clarke}, C.~J., {Tazzari}, M., {Juhasz}, A., {et~al.} 2018, \apjl, 866, L6

\bibitem[{{Claudi} {et~al.}(2019){Claudi}, {Maire}, {Mesa}, {Cheetham},
  {Fontanive}, {Gratton}, {Zurlo}, {Avenhaus}, {Bhowmik}, {Biller},
  {Boccaletti}, {Bonavita}, {Bonnefoy}, {Cascone}, {Chauvin}, {Delboulb{\'e}},
  {Desidera}, {D'Orazi}, {Feautrier}, {Feldt}, {Flammini Dotti}, {Girard},
  {Giro}, {Janson}, {Hagelberg}, {Keppler}, {Kopytova}, {Lacour}, {Lagrange},
  {Langlois}, {Lannier}, {Le Coroller}, {Menard}, {Messina}, {Meyer},
  {Millward}, {Olofsson}, {Pavlov}, {Peretti}, {Perrot}, {Pinte}, {Pragt},
  {Ramos}, {Rochat}, {Rodet}, {Roelfsema}, {Rouan}, {Salter}, {Schmidt},
  {Sissa}, {Thebault}, {Udry}, \& {Vigan}}]{Claudi2019}
{Claudi}, R., {Maire}, A.~L., {Mesa}, D., {et~al.} 2019, \aap, 622, A96

\bibitem[{{Close} {et~al.}(2014){Close}, {Follette}, {Males}, {Puglisi},
  {Xompero}, {Apai}, {Najita}, {Weinberger}, {Morzinski}, {Rodigas}, {Hinz},
  {Bailey}, \& {Briguglio}}]{Close2014}
{Close}, L.~M., {Follette}, K.~B., {Males}, J.~R., {et~al.} 2014, \apjl, 781,
  L30

\bibitem[{{Cugno} {et~al.}(2019){Cugno}, {Quanz}, {Hunziker}, {Stolker},
  {Schmid}, {Avenhaus}, {Baudoz}, {Bohn}, {Bonnefoy}, {Buenzli}, {Chauvin},
  {Cheetham}, {Desidera}, {Dominik}, {Feautrier}, {Feldt}, {Ginski}, {Girard},
  {Gratton}, {Hagelberg}, {Hugot}, {Janson}, {Lagrange}, {Langlois}, {Magnard},
  {Maire}, {Menard}, {Meyer}, {Milli}, {Mordasini}, {Pinte}, {Pragt},
  {Roelfsema}, {Rigal}, {Szul{\'a}gyi}, {van Boekel}, {van der Plas}, {Vigan},
  {Wahhaj}, \& {Zurlo}}]{cugno2019}
{Cugno}, G., {Quanz}, S.~P., {Hunziker}, S., {et~al.} 2019, \aap, 622, A156

\bibitem[{{Cumming} {et~al.}(2008){Cumming}, {Butler}, {Marcy}, {Vogt},
  {Wright}, \& {Fischer}}]{Cumming2008}
{Cumming}, A., {Butler}, R.~P., {Marcy}, G.~W., {et~al.} 2008, \pasp, 120, 531

\bibitem[{{Currie} {et~al.}(2017){Currie}, {Brittain}, {Grady}, {Kenyon}, \&
  {Muto}}]{currie2017}
{Currie}, T., {Brittain}, S., {Grady}, C.~A., {Kenyon}, S.~J., \& {Muto}, T.
  2017, Research Notes of the American Astronomical Society, 1, 40

\bibitem[{{Currie} {et~al.}(2015){Currie}, {Cloutier}, {Brittain}, {Grady},
  {Burrows}, {Muto}, {Kenyon}, \& {Kuchner}}]{Currie2015}
{Currie}, T., {Cloutier}, R., {Brittain}, S., {et~al.} 2015, \apjl, 814, L27

\bibitem[{{Currie} \& {Sicilia-Aguilar}(2011)}]{Currie2011}
{Currie}, T., \& {Sicilia-Aguilar}, A. 2011, \apj, 732, 24

\bibitem[{{Currie} {et~al.}(2019){Currie}, {Marois}, {Cieza}, {Mulders},
  {Lawson}, {Caceres}, {Rodriguez-Ruiz}, {Wisniewski}, {Guyon}, {Brandt},
  {Kasdin}, {Groff}, {Lozi}, {Chilcote}, {Hodapp}, {Jovanovic}, {Martinache},
  {Skaf}, {Lyra}, {Tamura}, {Asensio-Torres}, {Dong}, {Grady}, {Gerard},
  {Fukagawa}, {Hand}, {Hayashi}, {Henning}, {Kudo}, {Kuzuhara}, {Kwon},
  {McElwain}, \& {Uyama}}]{currie2019}
{Currie}, T., {Marois}, C., {Cieza}, L., {et~al.} 2019, \apjl, 877, L3

\bibitem[{{D'Angelo} {et~al.}(2010){D'Angelo}, {Durisen}, \&
  {Lissauer}}]{DAngelo2010}
{D'Angelo}, G., {Durisen}, R.~H., \& {Lissauer}, J.~J. 2010, {Giant Planet
  Formation}, ed. S.~{Seager}, 319--346

\bibitem[{{Dodson-Robinson} \& {Salyk}(2011)}]{dodson2011}
{Dodson-Robinson}, S.~E., \& {Salyk}, C. 2011, \apj, 738, 131

\bibitem[{{Dong} \& {Fung}(2017)}]{Dong2017spiral}
{Dong}, R., \& {Fung}, J. 2017, \apj, 835, 38

\bibitem[{{Dong} {et~al.}(2015{\natexlab{a}}){Dong}, {Hall}, {Rice}, \&
  {Chiang}}]{Dong2015GI}
{Dong}, R., {Hall}, C., {Rice}, K., \& {Chiang}, E. 2015{\natexlab{a}}, \apjl,
  812, L32

\bibitem[{{Dong} {et~al.}(2018{\natexlab{a}}){Dong}, {Najita}, \&
  {Brittain}}]{dong2018a}
{Dong}, R., {Najita}, J.~R., \& {Brittain}, S. 2018{\natexlab{a}}, \apj, 862,
  103

\bibitem[{{Dong} {et~al.}(2015{\natexlab{b}}){Dong}, {Zhu}, {Rafikov}, \&
  {Stone}}]{Dong2015a}
{Dong}, R., {Zhu}, Z., {Rafikov}, R.~R., \& {Stone}, J.~M. 2015{\natexlab{b}},
  \apjl, 809, L5

\bibitem[{{Dong} {et~al.}(2015{\natexlab{c}}){Dong}, {Zhu}, \&
  {Whitney}}]{Dong2015b}
{Dong}, R., {Zhu}, Z., \& {Whitney}, B. 2015{\natexlab{c}}, \apj, 809, 93

\bibitem[{{Dong} {et~al.}(2018{\natexlab{b}}){Dong}, {Liu}, {Eisner},
  {Andrews}, {Fung}, {Zhu}, {Chiang}, {Hashimoto}, {Liu}, {Casassus},
  {Esposito}, {Hasegawa}, {Muto}, {Pavlyuchenkov}, {Wilner}, {Akiyama},
  {Tamura}, \& {Wisniewski}}]{Dong2018b}
{Dong}, R., {Liu}, S.-y., {Eisner}, J., {et~al.} 2018{\natexlab{b}}, \apj, 860,
  124

\bibitem[{{Dupuy} {et~al.}(2019){Dupuy}, {Brandt}, {Kratter}, \&
  {Bowler}}]{Dupuy2019}
{Dupuy}, T.~J., {Brandt}, T.~D., {Kratter}, K.~M., \& {Bowler}, B.~P. 2019,
  \apjl, 871, L4

\bibitem[{{Eisner}(2015)}]{Eisner2015}
{Eisner}, J.~A. 2015, \apjl, 803, L4

\bibitem[{{Espaillat} {et~al.}(2014{\natexlab{a}}){Espaillat}, {Muzerolle},
  {Najita}, {Andrews}, {Zhu}, {Calvet}, {Kraus}, {Hashimoto}, {Kraus}, \&
  {D'Alessio}}]{Espaillat2014}
{Espaillat}, C., {Muzerolle}, J., {Najita}, J., {et~al.} 2014{\natexlab{a}},
  Protostars and Planets VI, 497

\bibitem[{{Espaillat} {et~al.}(2014{\natexlab{b}}){Espaillat}, {Muzerolle},
  {Najita}, {Andrews}, {Zhu}, {Calvet}, {Kraus}, {Hashimoto}, {Kraus}, \&
  {D'Alessio}}]{espaillat14}
{Espaillat}, C., {Muzerolle}, J., {Najita}, J., {et~al.} 2014{\natexlab{b}}, in
  Protostars and Planets VI, ed. H.~{Beuther}, R.~S. {Klessen}, C.~P.
  {Dullemond}, \& T.~{Henning}, 497

\bibitem[{{Fabrycky} \& {Murray-Clay}(2010)}]{Fab2010}
{Fabrycky}, D.~C., \& {Murray-Clay}, R.~A. 2010, \apj, 710, 1408

\bibitem[{{Fang} {et~al.}(2009){Fang}, {van Boekel}, {Wang}, {Carmona},
  {Sicilia-Aguilar}, \& {Henning}}]{Fang2009}
{Fang}, M., {van Boekel}, R., {Wang}, W., {et~al.} 2009, \aap, 504, 461

\bibitem[{{Follette} {et~al.}(2017){Follette}, {Rameau}, {Dong}, {Pueyo},
  {Close}, {Duch{\^e}ne}, {Fung}, {Leonard}, {Macintosh}, {Males}, {Marois},
  {Millar-Blanchaer}, {Morzinski}, {Mullen}, {Perrin}, {Spiro}, {Wang},
  {Ammons}, {Bailey}, {Barman}, {Bulger}, {Chilcote}, {Cotten}, {De Rosa},
  {Doyon}, {Fitzgerald}, {Goodsell}, {Graham}, {Greenbaum}, {Hibon}, {Hung},
  {Ingraham}, {Kalas}, {Konopacky}, {Larkin}, {Maire}, {Marchis}, {Metchev},
  {Nielsen}, {Oppenheimer}, {Palmer}, {Patience}, {Poyneer}, {Rajan},
  {Rantakyr{\"o}}, {Savransky}, {Schneider}, {Sivaramakrishnan}, {Song},
  {Soummer}, {Thomas}, {Vega}, {Wallace}, {Ward-Duong}, {Wiktorowicz}, \&
  {Wolff}}]{Follette2017}
{Follette}, K.~B., {Rameau}, J., {Dong}, R., {et~al.} 2017, \aj, 153, 264

\bibitem[{{Fortney} {et~al.}(2008){Fortney}, {Marley}, {Saumon}, \&
  {Lodders}}]{fortney2008}
{Fortney}, J.~J., {Marley}, M.~S., {Saumon}, D., \& {Lodders}, K. 2008, \apj,
  683, 1104

\bibitem[{{Fukagawa} {et~al.}(2006){Fukagawa}, {Tamura}, {Itoh}, {Kudo},
  {Imaeda}, {Oasa}, {Hayashi}, \& {Hayashi}}]{Fukagawa2006}
{Fukagawa}, M., {Tamura}, M., {Itoh}, Y., {et~al.} 2006, \apjl, 636, L153

\bibitem[{{Fukagawa} {et~al.}(2013){Fukagawa}, {Tsukagoshi}, {Momose}, {Saigo},
  {Ohashi}, {Kitamura}, {Inutsuka}, {Muto}, {Nomura}, {Takeuchi}, {Kobayashi},
  {Hanawa}, {Akiyama}, {Honda}, {Fujiwara}, {Kataoka}, {Takahashi}, \&
  {Shibai}}]{Fukagawa2013}
{Fukagawa}, M., {Tsukagoshi}, T., {Momose}, M., {et~al.} 2013, \pasj, 65, L14

\bibitem[{{Fung} {et~al.}(2015){Fung}, {Artymowicz}, \& {Wu}}]{Fung2015}
{Fung}, J., {Artymowicz}, P., \& {Wu}, Y. 2015, \apj, 811, 101

\bibitem[{{Fung} \& {Dong}(2015)}]{Fung2015spiral}
{Fung}, J., \& {Dong}, R. 2015, \apjl, 815, L21

\bibitem[{{Fung} {et~al.}(2014){Fung}, {Shi}, \& {Chiang}}]{Fung2014}
{Fung}, J., {Shi}, J.-M., \& {Chiang}, E. 2014, \apj, 782, 88

\bibitem[{{Furlan} {et~al.}(2011){Furlan}, {Luhman}, {Espaillat}, {D'Alessio},
  {Adame}, {Manoj}, {Kim}, {Watson}, {Forrest}, {McClure}, {Calvet}, {Sargent},
  {Green}, \& {Fischer}}]{Furlan2011}
{Furlan}, E., {Luhman}, K.~L., {Espaillat}, C., {et~al.} 2011, \apjs, 195, 3

\bibitem[{{Grady} {et~al.}(2001){Grady}, {Polomski}, {Henning}, {Stecklum},
  {Woodgate}, {Telesco}, {Pi{\~n}a}, {Gull}, {Boggess}, {Bowers}, {Bruhweiler},
  {Clampin}, {Danks}, {Green}, {Heap}, {Hutchings}, {Jenkins}, {Joseph},
  {Kaiser}, {Kimble}, {Kraemer}, {Lindler}, {Linsky}, {Maran}, {Moos}, {Plait},
  {Roesler}, {Timothy}, \& {Weistrop}}]{Grady2001}
{Grady}, C.~A., {Polomski}, E.~F., {Henning}, T., {et~al.} 2001, \aj, 122, 3396

\bibitem[{{Grady} {et~al.}(2013){Grady}, {Muto}, {Hashimoto}, {Fukagawa},
  {Currie}, {Biller}, {Thalmann}, {Sitko}, {Russell}, {Wisniewski}, {Dong},
  {Kwon}, {Sai}, {Hornbeck}, {Schneider}, {Hines}, {Moro Mart{\'\i}n}, {Feldt},
  {Henning}, {Pott}, {Bonnefoy}, {Bouwman}, {Lacour}, {Mueller}, {Juh{\'a}sz},
  {Crida}, {Chauvin}, {Andrews}, {Wilner}, {Kraus}, {Dahm}, {Robitaille},
  {Jang-Condell}, {Abe}, {Akiyama}, {Brandner}, {Brandt}, {Carson}, {Egner},
  {Follette}, {Goto}, {Guyon}, {Hayano}, {Hayashi}, {Hayashi}, {Hodapp},
  {Ishii}, {Iye}, {Janson}, {Kandori}, {Knapp}, {Kudo}, {Kusakabe}, {Kuzuhara},
  {Mayama}, {McElwain}, {Matsuo}, {Miyama}, {Morino}, {Nishimura}, {Pyo},
  {Serabyn}, {Suto}, {Suzuki}, {Takami}, {Takato}, {Terada}, {Tomono},
  {Turner}, {Watanabe}, {Yamada}, {Takami}, {Usuda}, \& {Tamura}}]{Grady2013}
{Grady}, C.~A., {Muto}, T., {Hashimoto}, J., {et~al.} 2013, \apj, 762, 48

\bibitem[{{Gratton} {et~al.}(2019){Gratton}, {Ligi}, {Sissa}, {Desidera},
  {Mesa}, {Bonnefoy}, {Chauvin}, {Cheetham}, {Feldt}, {Lagrange}, {Langlois},
  {Meyer}, {Vigan}, {Boccaletti}, {Janson}, {Lazzoni}, {Zurlo}, {De Boer},
  {Henning}, {D'Orazi}, {Gluck}, {Madec}, {Jaquet}, {Baudoz}, {Fantinel},
  {Pavlov}, \& {Wildi}}]{Gratton2019}
{Gratton}, R., {Ligi}, R., {Sissa}, E., {et~al.} 2019, \aap, 623, A140

\bibitem[{{Gregorio-Hetem} \& {Hetem}(2002)}]{Greg2002}
{Gregorio-Hetem}, J., \& {Hetem}, A. 2002, \mnras, 336, 197

\bibitem[{{Gressel} {et~al.}(2013){Gressel}, {Nelson}, {Turner}, \&
  {Ziegler}}]{Gressel2013}
{Gressel}, O., {Nelson}, R.~P., {Turner}, N.~J., \& {Ziegler}, U. 2013, \apj,
  779, 59

\bibitem[{{Guidi} {et~al.}(2018){Guidi}, {Ruane}, {Williams}, {Mawet}, {Testi},
  {Zurlo}, {Absil}, {Bottom}, {Choquet}, {Christiaens}, {Femen{\'\i}a
  Castell{\'a}}, {Huby}, {Isella}, {Kastner}, {Meshkat}, {Reggiani}, {Riggs},
  {Serabyn}, \& {Wallack}}]{Guidi2018}
{Guidi}, G., {Ruane}, G., {Williams}, J.~P., {et~al.} 2018, \mnras, 479, 1505

\bibitem[{{Haffert} {et~al.}(2019){Haffert}, {Bohn}, {de Boer}, {Snellen},
  {Brinchmann}, {Girard}, {Keller}, \& {Bacon}}]{haffert2019}
{Haffert}, S.~Y., {Bohn}, A.~J., {de Boer}, J., {et~al.} 2019, Nature
  Astronomy, 329

\bibitem[{{Hall} {et~al.}(2019){Hall}, {Dong}, {Rice}, {Harries}, {Najita},
  {Alexander}, \& {Brittain}}]{hall2019}
{Hall}, C., {Dong}, R., {Rice}, K., {et~al.} 2019, \apj, 871, 228

\bibitem[{{Hall} {et~al.}(2018){Hall}, {Rice}, {Dipierro}, {Forgan}, {Harries},
  \& {Alexander}}]{hall2018}
{Hall}, C., {Rice}, K., {Dipierro}, G., {et~al.} 2018, \mnras, 477, 1004

\bibitem[{{Hartmann} {et~al.}(1998){Hartmann}, {Calvet}, {Gullbring}, \&
  {D'Alessio}}]{Hartmann1998}
{Hartmann}, L., {Calvet}, N., {Gullbring}, E., \& {D'Alessio}, P. 1998, \apj,
  495, 385

\bibitem[{{Hartmann} {et~al.}(2016){Hartmann}, {Herczeg}, \&
  {Calvet}}]{Hartmann2016}
{Hartmann}, L., {Herczeg}, G., \& {Calvet}, N. 2016, \araa, 54, 135

\bibitem[{{Herbig}(1977)}]{Herbig1977}
{Herbig}, G.~H. 1977, \apj, 217, 693

\bibitem[{{Ingleby} {et~al.}(2013){Ingleby}, {Calvet}, {Herczeg}, {Blaty},
  {Walter}, {Ardila}, {Alexander}, {Edwards}, {Espaillat}, {Gregory},
  {Hillenbrand}, \& {Brown}}]{Ingleby2013}
{Ingleby}, L., {Calvet}, N., {Herczeg}, G., {et~al.} 2013, \apj, 767, 112

\bibitem[{{Isella} {et~al.}(2019){Isella}, {Benisty}, {Teague}, {Bae},
  {Keppler}, {Facchini}, \& {P{\'e}rez}}]{Isella2019}
{Isella}, A., {Benisty}, M., {Teague}, R., {et~al.} 2019, \apjl, 879, L25

\bibitem[{{Isella} {et~al.}(2014){Isella}, {Chandler}, {Carpenter},
  {P{\'e}rez}, \& {Ricci}}]{Isella2014}
{Isella}, A., {Chandler}, C.~J., {Carpenter}, J.~M., {P{\'e}rez}, L.~M., \&
  {Ricci}, L. 2014, \apj, 788, 129

\bibitem[{{Jensen} {et~al.}(2009){Jensen}, {Cohen}, \&
  {Gagn{\'e}}}]{Jensen2009}
{Jensen}, E. L.~N., {Cohen}, D.~H., \& {Gagn{\'e}}, M. 2009, \apj, 703, 252

\bibitem[{{Johnson} {et~al.}(2010){Johnson}, {Howard}, {Bowler}, {Henry},
  {Marcy}, {Wright}, {Fischer}, \& {Isaacson}}]{Johnson2010}
{Johnson}, J.~A., {Howard}, A.~W., {Bowler}, B.~P., {et~al.} 2010, \pasp, 122,
  701

\bibitem[{{Kenyon} \& {Hartmann}(1995)}]{Kenyon1995}
{Kenyon}, S.~J., \& {Hartmann}, L. 1995, \apjs, 101, 117

\bibitem[{{Kenyon} {et~al.}(1990){Kenyon}, {Hartmann}, {Strom}, \&
  {Strom}}]{kenyon1990}
{Kenyon}, S.~J., {Hartmann}, L.~W., {Strom}, K.~M., \& {Strom}, S.~E. 1990,
  \aj, 99, 869

\bibitem[{{Keppler} {et~al.}(2018){Keppler}, {Benisty}, {M{\"u}ller},
  {Henning}, {van Boekel}, {Cantalloube}, {Ginski}, {van Holstein}, {Maire},
  {Pohl}, {Samland}, {Avenhaus}, {Baudino}, {Boccaletti}, {de Boer},
  {Bonnefoy}, {Chauvin}, {Desidera}, {Langlois}, {Lazzoni}, {Marleau},
  {Mordasini}, {Pawellek}, {Stolker}, {Vigan}, {Zurlo}, {Birnstiel},
  {Brandner}, {Feldt}, {Flock}, {Girard}, {Gratton}, {Hagelberg}, {Isella},
  {Janson}, {Juhasz}, {Kemmer}, {Kral}, {Lagrange}, {Launhardt}, {Matter},
  {M{\'e}nard}, {Milli}, {Molli{\`e}re}, {Olofsson}, {P{\'e}rez}, {Pinilla},
  {Pinte}, {Quanz}, {Schmidt}, {Udry}, {Wahhaj}, {Williams}, {Buenzli},
  {Cudel}, {Dominik}, {Galicher}, {Kasper}, {Lannier}, {Mesa}, {Mouillet},
  {Peretti}, {Perrot}, {Salter}, {Sissa}, {Wildi}, {Abe}, {Antichi},
  {Augereau}, {Baruffolo}, {Baudoz}, {Bazzon}, {Beuzit}, {Blanchard}, {Brems},
  {Buey}, {De Caprio}, {Carbillet}, {Carle}, {Cascone}, {Cheetham}, {Claudi},
  {Costille}, {Delboulb{\'e}}, {Dohlen}, {Fantinel}, {Feautrier}, {Fusco},
  {Giro}, {Gluck}, {Gry}, {Hubin}, {Hugot}, {Jaquet}, {Le Mignant}, {Llored},
  {Madec}, {Magnard}, {Martinez}, {Maurel}, {Meyer}, {M{\"o}ller-Nilsson},
  {Moulin}, {Mugnier}, {Orign{\'e}}, {Pavlov}, {Perret}, {Petit}, {Pragt},
  {Puget}, {Rabou}, {Ramos}, {Rigal}, {Rochat}, {Roelfsema}, {Rousset}, {Roux},
  {Salasnich}, {Sauvage}, {Sevin}, {Soenke}, {Stadler}, {Suarez}, {Turatto}, \&
  {Weber}}]{Keppler2018}
{Keppler}, M., {Benisty}, M., {M{\"u}ller}, A., {et~al.} 2018, \aap, 617, A44

\bibitem[{{Klahr} \& {Kley}(2006)}]{Klahr2006}
{Klahr}, H., \& {Kley}, W. 2006, \aap, 445, 747

\bibitem[{{Kratter} \& {Lodato}(2016)}]{Kratter2016}
{Kratter}, K., \& {Lodato}, G. 2016, \araa, 54, 271

\bibitem[{{Kraus} \& {Ireland}(2012)}]{Kraus2012}
{Kraus}, A.~L., \& {Ireland}, M.~J. 2012, \apj, 745, 5

\bibitem[{{Kraus} {et~al.}(2013){Kraus}, {Ireland}, {Sitko}, {Monnier},
  {Calvet}, {Espaillat}, {Grady}, {Harries}, {H{\"o}nig}, {Russell},
  {Swearingen}, {Werren}, \& {Wilner}}]{Kraus2013}
{Kraus}, S., {Ireland}, M.~J., {Sitko}, M.~L., {et~al.} 2013, The Astrophysical
  Journal, 768, 80

\bibitem[{{Lacour} {et~al.}(2016){Lacour}, {Biller}, {Cheetham}, {Greenbaum},
  {Pearce}, {Marino}, {Tuthill}, {Pueyo}, {Mamajek}, {Girard},
  {Sivaramakrishnan}, {Bonnefoy}, {Baraffe}, {Chauvin}, {Olofsson}, {Juhasz},
  {Benisty}, {Pott}, {Sicilia-Aguilar}, {Henning}, {Cardwell}, {Goodsell},
  {Graham}, {Hibon}, {Ingraham}, {Konopacky}, {Macintosh}, {Oppenheimer},
  {Perrin}, {Rantakyr{\"o}}, {Sadakuni}, \& {Thomas}}]{Lacour2016}
{Lacour}, S., {Biller}, B., {Cheetham}, A., {et~al.} 2016, \aap, 590, A90

\bibitem[{{Lagrange} {et~al.}(2009){Lagrange}, {Gratadour}, {Chauvin}, {Fusco},
  {Ehrenreich}, {Mouillet}, {Rousset}, {Rouan}, {Allard}, {Gendron}, {Charton},
  {Mugnier}, {Rabou}, {Montri}, \& {Lacombe}}]{Lagrange2009}
{Lagrange}, A.-M., {Gratadour}, D., {Chauvin}, G., {et~al.} 2009, \aap, 493,
  L21

\bibitem[{{Ligi} {et~al.}(2018){Ligi}, {Vigan}, {Gratton}, {de Boer},
  {Benisty}, {Boccaletti}, {Quanz}, {Meyer}, {Ginski}, {Sissa}, {Gry},
  {Henning}, {Beuzit}, {Biller}, {Bonnefoy}, {Chauvin}, {Cheetham}, {Cudel},
  {Delorme}, {Desidera}, {Feldt}, {Galicher}, {Girard}, {Janson}, {Kasper},
  {Kopytova}, {Lagrange}, {Langlois}, {Lecoroller}, {Maire}, {M{\'e}nard},
  {Mesa}, {Peretti}, {Perrot}, {Pinilla}, {Pohl}, {Rouan}, {Stolker},
  {Samland}, {Wahhaj}, {Wildi}, {Zurlo}, {Buey}, {Fantinel}, {Fusco}, {Jaquet},
  {Moulin}, {Ramos}, {Suarez}, \& {Weber}}]{Ligi2018}
{Ligi}, R., {Vigan}, A., {Gratton}, R., {et~al.} 2018, \mnras, 473, 1774

\bibitem[{{Liskowsky} {et~al.}(2012){Liskowsky}, {Brittain}, {Najita}, {Carr},
  {Doppmann}, \& {Troutman}}]{Lisk2012}
{Liskowsky}, J.~P., {Brittain}, S.~D., {Najita}, J.~R., {et~al.} 2012, \apj,
  760, 153

\bibitem[{{Long} {et~al.}(2018){Long}, {Akiyama}, {Sitko}, {Fernandes},
  {Assani}, {Grady}, {Cure}, {Danchi}, {Dong}, {Fukagawa}, {Hasegawa},
  {Hashimoto}, {Henning}, {Inutsuka}, {Kraus}, {Kwon}, {Lisse}, {Baobabu Liu},
  {Mayama}, {Muto}, {Nakagawa}, {Takami}, {Tamura}, {Currie}, {Wisniewski}, \&
  {Yang}}]{Long2018}
{Long}, Z.~C., {Akiyama}, E., {Sitko}, M., {et~al.} 2018, \apj, 858, 112

\bibitem[{{Lubow} \& {D'Angelo}(2006)}]{Lubow2006}
{Lubow}, S.~H., \& {D'Angelo}, G. 2006, \apj, 641, 526

\bibitem[{{Lubow} \& {Martin}(2012)}]{Lubow2012}
{Lubow}, S.~H., \& {Martin}, R.~G. 2012, \apjl, 749, L37

\bibitem[{{Lubow} {et~al.}(1999){Lubow}, {Seibert}, \&
  {Artymowicz}}]{lubow1999}
{Lubow}, S.~H., {Seibert}, M., \& {Artymowicz}, P. 1999, \apj, 526, 1001

\bibitem[{{Luhn} {et~al.}(2018){Luhn}, {Bastien}, {Wright}, {Johnson},
  {Howard}, \& {Isaacson}}]{Luhn2018}
{Luhn}, J.~K., {Bastien}, F.~A., {Wright}, J.~T., {et~al.} 2018, ArXiv
  e-prints, arXiv:1811.03043

\bibitem[{{Maire} {et~al.}(2017){Maire}, {Stolker}, {Messina}, {M{\"u}ller},
  {Biller}, {Currie}, {Dominik}, {Grady}, {Boccaletti}, {Bonnefoy}, {Chauvin},
  {Galicher}, {Millward}, {Pohl}, {Brandner}, {Henning}, {Lagrange},
  {Langlois}, {Meyer}, {Quanz}, {Vigan}, {Zurlo}, {van Boekel}, {Buenzli},
  {Buey}, {Desidera}, {Feldt}, {Fusco}, {Ginski}, {Giro}, {Gratton}, {Hubin},
  {Lannier}, {Le Mignant}, {Mesa}, {Peretti}, {Perrot}, {Ramos}, {Salter},
  {Samland}, {Sissa}, {Stadler}, {Thalmann}, {Udry}, \& {Weber}}]{maire2017}
{Maire}, A.~L., {Stolker}, T., {Messina}, S., {et~al.} 2017, \aap, 601, A134

\bibitem[{{Malfait} {et~al.}(1998){Malfait}, {Bogaert}, \&
  {Waelkens}}]{Malfait1998}
{Malfait}, K., {Bogaert}, E., \& {Waelkens}, C. 1998, \aap, 331, 211

\bibitem[{{Marino} {et~al.}(2015){Marino}, {Perez}, \& {Casassus}}]{Marino2015}
{Marino}, S., {Perez}, S., \& {Casassus}, S. 2015, \apjl, 798, L44

\bibitem[{{Marley} {et~al.}(2007){Marley}, {Fortney}, {Hubickyj},
  {Bodenheimer}, \& {Lissauer}}]{marley2007}
{Marley}, M.~S., {Fortney}, J.~J., {Hubickyj}, O., {Bodenheimer}, P., \&
  {Lissauer}, J.~J. 2007, \apj, 655, 541

\bibitem[{{Marois} {et~al.}(2008){Marois}, {Macintosh}, {Barman}, {Zuckerman},
  {Song}, {Patience}, {Lafreni{\`e}re}, \& {Doyon}}]{Marois2008}
{Marois}, C., {Macintosh}, B., {Barman}, T., {et~al.} 2008, Science, 322, 1348

\bibitem[{{Martin} \& {Lubow}(2011)}]{martin2011}
{Martin}, R.~G., \& {Lubow}, S.~H. 2011, \apjl, 740, L6

\bibitem[{{Mawet} {et~al.}(2017){Mawet}, {Choquet}, {Absil}, {Huby}, {Bottom},
  {Serabyn}, {Femenia}, {Lebreton}, {Matthews}, {Gomez Gonzalez}, {Wertz},
  {Carlomagno}, {Christiaens}, {Defr{\`e}re}, {Delacroix}, {Forsberg},
  {Habraken}, {Jolivet}, {Karlsson}, {Milli}, {Pinte}, {Piron}, {Reggiani},
  {Surdej}, \& {Vargas Catalan}}]{Mawet2017}
{Mawet}, D., {Choquet}, {\'E}., {Absil}, O., {et~al.} 2017, \aj, 153, 44

\bibitem[{{Mendigut{\'{\i}}a} {et~al.}(2014){Mendigut{\'{\i}}a}, {Fairlamb},
  {Montesinos}, {Oudmaijer}, {Najita}, {Brittain}, \& {van den
  Ancker}}]{Mend2014}
{Mendigut{\'{\i}}a}, I., {Fairlamb}, J., {Montesinos}, B., {et~al.} 2014, \apj,
  790, 21

\bibitem[{{Mendigut{\'{\i}}a} {et~al.}(2018){Mendigut{\'{\i}}a}, {Oudmaijer},
  {Schneider}, {Hu{\'e}lamo}, {Baines}, {Brittain}, \& {Aberasturi}}]{Mend2018}
{Mendigut{\'{\i}}a}, I., {Oudmaijer}, R.~D., {Schneider}, P.~C., {et~al.} 2018,
  \aap, 618, L9

\bibitem[{{Muto} {et~al.}(2012){Muto}, {Grady}, {Hashimoto}, {Fukagawa},
  {Hornbeck}, {Sitko}, {Russell}, {Werren}, {Cur{\'e}}, {Currie}, {Ohashi},
  {Okamoto}, {Momose}, {Honda}, {Inutsuka}, {Takeuchi}, {Dong}, {Abe},
  {Brandner}, {Brandt}, {Carson}, {Egner}, {Feldt}, {Fukue}, {Goto}, {Guyon},
  {Hayano}, {Hayashi}, {Hayashi}, {Henning}, {Hodapp}, {Ishii}, {Iye},
  {Janson}, {Kandori}, {Knapp}, {Kudo}, {Kusakabe}, {Kuzuhara}, {Matsuo},
  {Mayama}, {McElwain}, {Miyama}, {Morino}, {Moro-Martin}, {Nishimura}, {Pyo},
  {Serabyn}, {Suto}, {Suzuki}, {Takami}, {Takato}, {Terada}, {Thalmann},
  {Tomono}, {Turner}, {Watanabe}, {Wisniewski}, {Yamada}, {Takami}, {Usuda}, \&
  {Tamura}}]{Muto2012}
{Muto}, T., {Grady}, C.~A., {Hashimoto}, J., {et~al.} 2012, \apjl, 748, L22

\bibitem[{{Muzerolle} {et~al.}(2010{\natexlab{a}}){Muzerolle}, {Allen},
  {Megeath}, {Hern{\'a}ndez}, \& {Gutermuth}}]{muz2010}
{Muzerolle}, J., {Allen}, L.~E., {Megeath}, S.~T., {Hern{\'a}ndez}, J., \&
  {Gutermuth}, R.~A. 2010{\natexlab{a}}, \apj, 708, 1107

\bibitem[{{Muzerolle} {et~al.}(2010{\natexlab{b}}){Muzerolle}, {Allen},
  {Megeath}, {Hern{\'a}ndez}, \& {Gutermuth}}]{muzerolle2010}
---. 2010{\natexlab{b}}, \apj, 708, 1107

\bibitem[{{Nielsen} {et~al.}(2019){Nielsen}, {De Rosa}, {Macintosh}, {Wang},
  {Ruffio}, {Chiang}, {Marley}, {Saumon}, {Savransky}, {Ammons}, {Bailey},
  {Barman}, {Blain}, {Bulger}, {Burrows}, {Chilcote}, {Cotten}, {Czekala},
  {Doyon}, {Duch{\^e}ne}, {Esposito}, {Fabrycky}, {Fitzgerald}, {Follette},
  {Fortney}, {Gerard}, {Goodsell}, {Graham}, {Greenbaum}, {Hibon}, {Hinkley},
  {Hirsch}, {Hom}, {Hung}, {Dawson}, {Ingraham}, {Kalas}, {Konopacky},
  {Larkin}, {Lee}, {Lin}, {Maire}, {Marchis}, {Marois}, {Metchev},
  {Millar-Blanchaer}, {Morzinski}, {Oppenheimer}, {Palmer}, {Patience},
  {Perrin}, {Poyneer}, {Pueyo}, {Rafikov}, {Rajan}, {Rameau}, {Rantakyr{\"o}},
  {Ren}, {Schneider}, {Sivaramakrishnan}, {Song}, {Soummer}, {Tallis},
  {Thomas}, {Ward-Duong}, \& {Wolff}}]{Nielsen2019}
{Nielsen}, E.~L., {De Rosa}, R.~J., {Macintosh}, B., {et~al.} 2019, \aj, 158,
  13

\bibitem[{{P{\'e}rez} {et~al.}(2019){P{\'e}rez}, {Casassus}, {Hales}, {Marino},
  {Cheetham}, {Zurlo}, {Cieza}, {Dong}, {Alarc{\'o}n}, {Ben{\'\i}tez-Llambay},
  \& {Fomalont}}]{Perez2019}
{P{\'e}rez}, S., {Casassus}, S., {Hales}, A., {et~al.} 2019, arXiv e-prints,
  arXiv:1906.06305

\bibitem[{{Pineda} {et~al.}(2019){Pineda}, {Szul{\'a}gyi}, {Quanz}, {van
  Dishoeck}, {Garufi}, {Meru}, {Mulders}, {Testi}, {Meyer}, \&
  {Reggiani}}]{Pineda2019}
{Pineda}, J.~E., {Szul{\'a}gyi}, J., {Quanz}, S.~P., {et~al.} 2019, \apj, 871,
  48

\bibitem[{{Pontoppidan} {et~al.}(2011){Pontoppidan}, {Blake}, \&
  {Smette}}]{Pont2011}
{Pontoppidan}, K.~M., {Blake}, G.~A., \& {Smette}, A. 2011, \apj, 733, 84

\bibitem[{{Price} {et~al.}(2018){Price}, {Cuello}, {Pinte}, {Mentiplay},
  {Casassus}, {Christiaens}, {Kennedy}, {Cuadra}, {Sebastian Perez}, {Marino},
  {Armitage}, {Zurlo}, {Juhasz}, {Ragusa}, {Laibe}, \& {Lodato}}]{Price2018}
{Price}, D.~J., {Cuello}, N., {Pinte}, C., {et~al.} 2018, \mnras, 477, 1270

\bibitem[{{Quanz} {et~al.}(2015){Quanz}, {Amara}, {Meyer}, {Girard},
  {Kenworthy}, \& {Kasper}}]{quanz2015ApJ}
{Quanz}, S.~P., {Amara}, A., {Meyer}, M.~R., {et~al.} 2015, \apj, 807, 64

\bibitem[{{Quanz} {et~al.}(2013){Quanz}, {Amara}, {Meyer}, {Kenworthy},
  {Kasper}, \& {Girard}}]{quanz2013}
---. 2013, \apjl, 766, L1

\bibitem[{{Rameau} {et~al.}(2017){Rameau}, {Follette}, {Pueyo}, {Marois},
  {Macintosh}, {Millar-Blanchaer}, {Wang}, {Vega}, {Doyon}, {Lafreni{\`e}re},
  {Nielsen}, {Bailey}, {Chilcote}, {Close}, {Esposito}, {Males}, {Metchev},
  {Morzinski}, {Ruffio}, {Wolff}, {Ammons}, {Barman}, {Bulger}, {Cotten}, {De
  Rosa}, {Duchene}, {Fitzgerald}, {Goodsell}, {Graham}, {Greenbaum}, {Hibon},
  {Hung}, {Ingraham}, {Kalas}, {Konopacky}, {Larkin}, {Maire}, {Marchis},
  {Oppenheimer}, {Palmer}, {Patience}, {Perrin}, {Poyneer}, {Rajan},
  {Rantakyr{\"o}}, {Marley}, {Savransky}, {Schneider}, {Sivaramakrishnan},
  {Song}, {Soummer}, {Thomas}, {Wallace}, {Ward-Duong}, \&
  {Wiktorowicz}}]{Rameau2017}
{Rameau}, J., {Follette}, K.~B., {Pueyo}, L., {et~al.} 2017, \aj, 153, 244

\bibitem[{{Reggiani} {et~al.}(2018){Reggiani}, {Christiaens}, {Absil}, {Mawet},
  {Huby}, {Choquet}, {Gomez Gonzalez}, {Ruane}, {Femenia}, {Serabyn},
  {Matthews}, {Barraza}, {Carlomagno}, {Defr{\`e}re}, {Delacroix}, {Habraken},
  {Jolivet}, {Karlsson}, {Orban de Xivry}, {Piron}, {Surdej}, {Vargas Catalan},
  \& {Wertz}}]{Reggiani2018k}
{Reggiani}, M., {Christiaens}, V., {Absil}, O., {et~al.} 2018, \aap, 611, A74

\bibitem[{{Rice} {et~al.}(2006){Rice}, {Armitage}, {Wood}, \&
  {Lodato}}]{rice2006}
{Rice}, W.~K.~M., {Armitage}, P.~J., {Wood}, K., \& {Lodato}, G. 2006, \mnras,
  373, 1619

\bibitem[{{Sallum} {et~al.}(2015){Sallum}, {Follette}, {Eisner}, {Close},
  {Hinz}, {Kratter}, {Males}, {Skemer}, {Macintosh}, {Tuthill}, {Bailey},
  {Defr{\`e}re}, {Morzinski}, {Rodigas}, {Spalding}, {Vaz}, \&
  {Weinberger}}]{Sallum2015}
{Sallum}, S., {Follette}, K.~B., {Eisner}, J.~A., {et~al.} 2015, \nat, 527, 342

\bibitem[{{Sicilia-Aguilar} {et~al.}(2005){Sicilia-Aguilar}, {Hartmann},
  {Hern{\'a}ndez}, {Brice{\~n}o}, \& {Calvet}}]{Sicilia2005}
{Sicilia-Aguilar}, A., {Hartmann}, L.~W., {Hern{\'a}ndez}, J., {Brice{\~n}o},
  C., \& {Calvet}, N. 2005, \aj, 130, 188

\bibitem[{{Sissa} {et~al.}(2018){Sissa}, {Gratton}, {Garufi}, {Rigliaco},
  {Zurlo}, {Mesa}, {Langlois}, {de Boer}, {Desidera}, {Ginski}, {Lagrange},
  {Maire}, {Vigan}, {Dima}, {Antichi}, {Baruffolo}, {Bazzon}, {Benisty},
  {Beuzit}, {Biller}, {Boccaletti}, {Bonavita}, {Bonnefoy}, {Brandner},
  {Bruno}, {Buenzli}, {Cascone}, {Chauvin}, {Cheetham}, {Claudi}, {Cudel}, {De
  Caprio}, {Dominik}, {Fantinel}, {Farisato}, {Feldt}, {Fontanive}, {Galicher},
  {Giro}, {Hagelberg}, {Incorvaia}, {Janson}, {Kasper}, {Keppler}, {Kopytova},
  {Lagadec}, {Lannier}, {Lazzoni}, {LeCoroller}, {Lessio}, {Ligi}, {Marzari},
  {Menard}, {Meyer}, {Mouillet}, {Peretti}, {Perrot}, {Potiron}, {Rouan},
  {Salasnich}, {Salter}, {Samland}, {Schmidt}, {Scuderi}, \&
  {Wildi}}]{Sissa2018}
{Sissa}, E., {Gratton}, R., {Garufi}, A., {et~al.} 2018, \aap, 619, A160

\bibitem[{{Snellen} \& {Brown}(2018)}]{Snellen2018}
{Snellen}, I.~A.~G., \& {Brown}, A.~G.~A. 2018, Nature Astronomy, 2, 883

\bibitem[{{Stone} {et~al.}(2018){Stone}, {Skemer}, {Hinz}, {Bonavita},
  {Kratter}, {Maire}, {Defrere}, {Bailey}, {Spalding}, {Leisenring},
  {Desidera}, {Bonnefoy}, {Biller}, {Woodward}, {Henning}, {Skrutskie},
  {Eisner}, {Crepp}, {Patience}, {Weigelt}, {De Rosa}, {Schlieder}, {Brandner},
  {Apai}, {Su}, {Ertel}, {Ward-Duong}, {Morzinski}, {Schertl}, {Hofmann},
  {Close}, {Brems}, {Fortney}, {Oza}, {Buenzli}, \& {Bass}}]{Stone2018}
{Stone}, J.~M., {Skemer}, A.~J., {Hinz}, P.~M., {et~al.} 2018, ArXiv e-prints,
  arXiv:1810.10560

\bibitem[{{Szul{\'a}gyi}(2017)}]{szulagyi2017-temp}
{Szul{\'a}gyi}, J. 2017, \apj, 842, 103

\bibitem[{{Szul{\'a}gyi} {et~al.}(2019){Szul{\'a}gyi}, {Dullemond}, {Pohl}, \&
  {Quanz}}]{Szulagyi2019}
{Szul{\'a}gyi}, J., {Dullemond}, C.~P., {Pohl}, A., \& {Quanz}, S.~P. 2019,
  \mnras, 487, 1248

\bibitem[{{Szul{\'a}gyi} {et~al.}(2016){Szul{\'a}gyi}, {Masset}, {Lega},
  {Crida}, {Morbidelli}, \& {Guillot}}]{szulagyi2016-disc}
{Szul{\'a}gyi}, J., {Masset}, F., {Lega}, E., {et~al.} 2016, \mnras, 460, 2853

\bibitem[{{Szul{\'a}gyi} {et~al.}(2014){Szul{\'a}gyi}, {Morbidelli}, {Crida},
  \& {Masset}}]{szulagyi2014-acc}
{Szul{\'a}gyi}, J., {Morbidelli}, A., {Crida}, A., \& {Masset}, F. 2014, \apj,
  782, 65

\bibitem[{{Szul{\'a}gyi} \& {Mordasini}(2017)}]{szulagyi2017-thermo}
{Szul{\'a}gyi}, J., \& {Mordasini}, C. 2017, \mnras, 465, L64

\bibitem[{{Tang} {et~al.}(2017){Tang}, {Guilloteau}, {Dutrey}, {Muto}, {Shen},
  {Gu}, {Inutsuka}, {Momose}, {Pietu}, {Fukagawa}, {Chapillon}, {Ho}, {di
  Folco}, {Corder}, {Ohashi}, \& {Hashimoto}}]{Tang2017}
{Tang}, Y.-W., {Guilloteau}, S., {Dutrey}, A., {et~al.} 2017, \apj, 840, 32

\bibitem[{{Tanigawa} {et~al.}(2012){Tanigawa}, {Ohtsuki}, \&
  {Machida}}]{Tanigawa2012}
{Tanigawa}, T., {Ohtsuki}, K., \& {Machida}, M.~N. 2012, \apj, 747, 47

\bibitem[{{Testi} {et~al.}(2015){Testi}, {Skemer}, {Henning}, {Bailey},
  {Defr{\`e}re}, {Hinz}, {Leisenring}, {Vaz}, {Esposito}, {Fontana}, {Marconi},
  {Skrutskie}, \& {Veillet}}]{Testi2015}
{Testi}, L., {Skemer}, A., {Henning}, T., {et~al.} 2015, \apj, 812, L38

\bibitem[{{Thalmann} {et~al.}(2016){Thalmann}, {Janson}, {Garufi},
  {Boccaletti}, {Quanz}, {Sissa}, {Gratton}, {Salter}, {Benisty}, {Bonnefoy},
  {Chauvin}, {Daemgen}, {Desidera}, {Dominik}, {Engler}, {Feldt}, {Henning},
  {Lagrange}, {Langlois}, {Lannier}, {Le Coroller}, {Ligi}, {M{\'e}nard},
  {Mesa}, {Meyer}, {Mulders}, {Olofsson}, {Pinte}, {Schmid}, {Vigan}, \&
  {Zurlo}}]{Thalmann2016}
{Thalmann}, C., {Janson}, M., {Garufi}, A., {et~al.} 2016, \apjl, 828, L17

\bibitem[{{Uyama} {et~al.}(2017){Uyama}, {Hashimoto}, {Kuzuhara}, {Mayama},
  {Akiyama}, {Currie}, {Livingston}, {Kudo}, {Kusakabe}, {Abe}, {Brandner},
  {Brandt}, {Carson}, {Egner}, {Feldt}, {Goto}, {Grady}, {Guyon}, {Hayano},
  {Hayashi}, {Hayashi}, {Henning}, {Hodapp}, {Ishii}, {Iye}, {Janson},
  {Kandori}, {Knapp}, {Kwon}, {Matsuo}, {Mcelwain}, {Miyama}, {Morino},
  {Moro-Martin}, {Nishimura}, {Pyo}, {Serabyn}, {Suenaga}, {Suto}, {Suzuki},
  {Takahashi}, {Takami}, {Takato}, {Terada}, {Thalmann}, {Turner}, {Watanabe},
  {Wisniewski}, {Yamada}, {Takami}, {Usuda}, \& {Tamura}}]{Uyama2017}
{Uyama}, T., {Hashimoto}, J., {Kuzuhara}, M., {et~al.} 2017, \aj, 153, 106

\bibitem[{{van der Marel} {et~al.}(2016){van der Marel}, {Verhaar}, {van
  Terwisga}, {Mer{\'\i}n}, {Herczeg}, {Ligterink}, \& {van
  Dishoeck}}]{vanderMarel2016}
{van der Marel}, N., {Verhaar}, B.~W., {van Terwisga}, S., {et~al.} 2016, \aap,
  592, A126

\bibitem[{{Wagner} {et~al.}(2019){Wagner}, {Stone}, {Spalding}, {Apai}, {Dong},
  {Ertel}, {Leisenring}, \& {Webster}}]{Wagner2019}
{Wagner}, K., {Stone}, J.~M., {Spalding}, E., {et~al.} 2019, \apj, 882, 20

\bibitem[{{Wagner} {et~al.}(2018){Wagner}, {Follete}, {Close}, {Apai}, {Gibbs},
  {Keppler}, {M{\"u}ller}, {Henning}, {Kasper}, {Wu}, {Long}, {Males},
  {Morzinski}, \& {McClure}}]{Wagner2018}
{Wagner}, K., {Follete}, K.~B., {Close}, L.~M., {et~al.} 2018, \apjl, 863, L8

\bibitem[{{Wang} {et~al.}(2018){Wang}, {Graham}, {Dawson}, {Fabrycky}, {De
  Rosa}, {Pueyo}, {Konopacky}, {Macintosh}, {Marois}, {Chiang}, {Ammons},
  {Arriaga}, {Bailey}, {Barman}, {Bulger}, {Chilcote}, {Cotten}, {Doyon},
  {Duch{\^e}ne}, {Esposito}, {Fitzgerald}, {Follette}, {Gerard}, {Goodsell},
  {Greenbaum}, {Hibon}, {Hung}, {Ingraham}, {Kalas}, {Larkin}, {Maire},
  {Marchis}, {Marley}, {Metchev}, {Millar-Blanchaer}, {Nielsen}, {Oppenheimer},
  {Palmer}, {Patience}, {Perrin}, {Poyneer}, {Rajan}, {Rameau},
  {Rantakyr{\"o}}, {Ruffio}, {Savransky}, {Schneider}, {Sivaramakrishnan},
  {Song}, {Soummer}, {Thomas}, {Wallace}, {Ward-Duong}, {Wiktorowicz}, \&
  {Wolff}}]{wang2018}
{Wang}, J.~J., {Graham}, J.~R., {Dawson}, R., {et~al.} 2018, \aj, 156, 192

\bibitem[{{Willson} {et~al.}(2016){Willson}, {Kraus}, {Kluska}, {Monnier},
  {Ireland}, {Aarnio}, {Sitko}, {Calvet}, {Espaillat}, \&
  {Wilner}}]{Willson2016}
{Willson}, M., {Kraus}, S., {Kluska}, J., {et~al.} 2016, \aap, 595, A9

\bibitem[{{Zhang} {et~al.}(2016){Zhang}, {Bergin}, {Blake}, {Cleeves},
  {Hogerheijde}, {Salinas}, \& {Schwarz}}]{Zhang2016}
{Zhang}, K., {Bergin}, E.~A., {Blake}, G.~A., {et~al.} 2016, \apjl, 818, L16

\bibitem[{{Zhu}(2015)}]{Zhu2015}
{Zhu}, Z. 2015, \apj, 799, 16

\bibitem[{{Zhu} {et~al.}(2018){Zhu}, {Andrews}, \& {Isella}}]{zhu2018}
{Zhu}, Z., {Andrews}, S.~M., \& {Isella}, A. 2018, \mnras, 479, 1850

\bibitem[{{Zhu} {et~al.}(2009){Zhu}, {Hartmann}, \& {Gammie}}]{Zhu2009}
{Zhu}, Z., {Hartmann}, L., \& {Gammie}, C. 2009, \apj, 694, 1045

\bibitem[{{Zhu} {et~al.}(2016){Zhu}, {Ju}, \& {Stone}}]{Zhu2016}
{Zhu}, Z., {Ju}, W., \& {Stone}, J.~M. 2016, \apj, 832, 193

\bibitem[{{Zhu} {et~al.}(2011){Zhu}, {Nelson}, {Hartmann}, {Espaillat}, \&
  {Calvet}}]{zhu2011}
{Zhu}, Z., {Nelson}, R.~P., {Hartmann}, L., {Espaillat}, C., \& {Calvet}, N.
  2011, \apj, 729, 47

\bibitem[{{Zurlo} {et~al.}(2020){Zurlo}, {Cugno}, {Montesinos}, {Perez},
  {Canovas}, {Casassus}, {Christiaens}, {Cieza}, \& {Huelamo}}]{zurlo2020}
{Zurlo}, A., {Cugno}, G., {Montesinos}, M., {et~al.} 2020, \aap, 633, A119

\end{thebibliography}

\clearpage
\begin{turnpage}

\begin{deluxetable}{lccccccccccccccc}
\tabletypesize{\tiny}
\tablehead{ \colhead{Star}	&	\colhead{Type}	&	\colhead{Distance}	&	\colhead{Morphology}\tablenotemark{a} &	\colhead{Cavity} &	\colhead{Inner} & \colhead{Outer} &\multicolumn{3}{c}{Stellar} & \multicolumn{3}{c}{Contrast} &	\multicolumn{3}{c}{$\lambda$L$_{\lambda}$}			\\
\colhead{[1]}	&	\colhead{[2]}	&	\colhead{[3]}	&	\colhead{[4]} &	\colhead{[5]} &	\colhead{[6]} & \colhead{[7]} &	\colhead{[8]}&	\colhead{[9]}&	\colhead{[10]} &	\colhead{[11]}&	\colhead{[12]}&	\colhead{[13]} &	\colhead{[14]}&	\colhead{[15]}&	\colhead{[16]}			\\
\colhead{}  & \colhead{} & \colhead{}   & \colhead{} & \colhead{Size} & \colhead{Extent} & \colhead{Extent} & \colhead{H}	&	\colhead{K$\rm_s$}	&	\colhead{L}		&	\colhead{$\Delta$ H}	&	\colhead{$\Delta$K$\rm_s$}	&	\colhead{$\Delta$L}	&	\colhead{H}	&	\colhead{K$\rm_s$}	&	\colhead{L} \\
\colhead{}	& \colhead{} & \colhead{pc}	& \colhead{} & \colhead{au}   & \colhead{au}     & \colhead{au}     & \multicolumn{3}{c}{Magnitude}                             &\multicolumn{3}{c}{Magnitude}                                                               &  \multicolumn{3}{c}{(erg s$^{-1}$)}		}

\startdata	
AB Aur$^{1}$	       & IMS &	163	$\pm$9			&	T$^{25}$, MSp$^{26}$	&	70	&	32.6	&	489.0	&			5.062			&			4.23			&			3.254			&			--	        		&			2.09E-03			&			--	        		&				--	        	    	&		$\leq$	1.27E+32		&			--	\\
DM Tau$^{1, 2}$	       & LMS &	145	$\pm$7			&	T$^{25}$	            &	19	&	5.8	    &	435.0	&			9.757			&			9.522			&			9.458			&			1.00E-02			&			--	        		&			--	        		&		$\leq$		6.02E+30			&			--	            	&			--	\\
GM Aur$^{1}$	       & LMS &	160	$\pm$13			&	T$^{25}$	            &	28	&	16.0	&	480.0	&			8.603			&			8.283			&			8.369			&			7.59E-04			&			--	        		&			--	        		&		$\leq$		1.60E+30			&			--	            	&			--	\\
LkH$\alpha$ 330$^{2}$  & IMS &	311	$\pm$24			&	T$^{25}$, 2Sp$^{27}$	&	68	&		    &	62.2	&			7.917			&			7.03			&			6.072			&			--	        		&			5.75E-03			&			--	        		&				--	        	    	&		$\leq$	9.68E+31		&			--	\\
HD100546$^{3-7}$	   & IMS &	110	$\pm$6			&	T$^{28}$, MSp$^{29}$	&	13	&	11.0	&	88.0	&			5.962			&			5.418			&			4.201			&			2.51E-04			&			--	        		&			--	        		&		$\leq$		2.86E+30			&			--	            	&			--	\\
MWC 758$^{3, 8, 9}$	   & IMS &	160	$\pm$11			&	T$^{25}$, 2Sp$^{30}$	&	73	&	16.0	&	176.0	&			6.56			&			5.804			&			4.599			&			--	        		&			1.00E-04			&			1.58E-03			&				--	        	    	&		$\leq$	1.38E+30		&		$\leq$	1.79E+31	\\
SAO 206462$^{3,10}$	   & IMS &	136	$\pm$10			&	T$^{25}$, 2Sp$^{31}$	&	46	&	13.6	&	272.0	&			6.587			&			5.843			&			5.046			&			2.00E-05			&			1.00E-04			&			--	        		&		$\leq$		1.95E+29			&		$\leq$9.57E+29	    	&			--	\\
HD169142$^{3, 11, 12}$ & IMS &	114	$\pm$7			&	T$^{25}$, MSp$^{32}$	&	20	&	11.4	&	114.0	&			6.911			&			6.41			&			5.995			&			1.32E-04			&			1.32E-04			&			--	        		&		$\leq$		6.73E+29			&		$\leq$	5.28E+29		&			--	\\
TW Hya$^{2, 3}$	       & LMS &	60.1$\pm$2.5		&	T$^{25}$	            &	4	&	1.2	    &	12.0	&			7.558			&			7.297			&			7.101			&			--	        		&			6.31E-03			&			--	        		&				--	        	    	&		$\leq$3.10E+30	    	&			--	\\
HD 141569$^{13}$	   & IMS &  111$\pm$1	        &	T$^{33}$	            &	20	&	11.1	&	276.6	&		    --              &		    --              &		    --              &		    --                  &		   --                   &		    --                  &		        --                      &	$\leq$30M$\rm_J$ at L-band	&           -- \\
HD 142527$^{14}$	   & IMS &    157$\pm$1	        &	T$^{25}$	            &	140	&	31.4	&	471.0	&	   	    5.715	    	&		    4.980	    	&	   	4.280	    	    &			--	        		&			--	        		&			--	        		&			4.39E+32	        	&			2.35E+32    		&			8.12E+31	\\
PDS 70b$^{15,16}$	   & LMS & 113$\pm$5	        &	T$^{25}$	            &	60	&	17.0	&	113.0	&			8.823			&			8.542			&			8.026			&			--	        		&			6.31E-04			&			1.74E-03			&				--	        		&			3.51E+29	    	&			4.18E+29	\\
PDS 70c$^{16}$	       &     &                      &		                    &	--  &	--	    &	--		&			8.823			&			8.542			&			8.026			&			--	        		&			3.02E-04			&			2.29E-03			&				--	        		&			1.68E+29	    	&			5.46E+29	\\
LkCa 15b$^{1,17,18}$.  & LMS &	 159$\pm$8	        &	T$^{25}$	            &	50	&	8.0	    &	31.8	&			8.600			&			8.163			&			7.49			&			5.25E-03			&			3.98E-03			&			1.00E-02			&				1.10E+31			&			6.16E+30    		&			7.72E+30	\\
LkCa 15c$^{18}$	       &	  &	                    &		                    &	--	&	--	  	&	--		&			8.600			&			8.163			&			7.49			&			5.25E-03			&			3.98E-03			&			1.00E-02			&			9.70E+30	        	&			6.16E+30			&		    --	\\
FL Cha (T35)$^{19}$	   & LMS &	188	$\pm$	10		&	T$^{34}$	            &	15	&	5.6	    &	56.4	&			9.904			&			9.109			&			8.294			&			--	        		&			1.20E-02			&			--	        		&				--	        	    &		$\leq$	1.09E+31		&			--	\\
FP Tau$^{2}$	       & LMS &	128	$\pm$	7		&	T$^{35}$	            &	--	&	6.4	    & 25.6      &           9.175           &			8.873			&			8.38			&			--	        		&			1.74E-02			&			2.51E-02			&				--	        	    &		$\leq$	9.14E+30		&		    $\leq$	5.59E+30	\\
DZ Cha$^{20}$	       & LMS &	100.83$\pm$0.26	    &	T, 2Sp$^{27}$	        &	7	&	4.0	& 80.7      &	       --               &		    --              &		     --                 &	$\rm \Delta J$=8.5	        &		     --                 &		    --                  &		           --               &		    --                  &	            --             \\
RX J1604.3-2130A$^{1,21}$ &	LMS	& 150	$\pm$	8	&	T$^{25}$	            &	70	& 15.0	& 450.0	    &			9.103			&			8.506			&			7.548			&			1.58E-05			&			1.58E-05			&			--	        		&		$\leq$		1.86E+28			&		$\leq$	1.60E+28		&			--	\\
RXJ1615.3-3255$^{2}$   & LMS & 158	$\pm$	6		&	T$^{25}$	            &	30	&	3.2	&	31.6	&			8.777			&			8.558			&			8.528			&			--	        		&			1.10E-02			&			--	        		&				--	        	    	&		$\leq$	1.16E+31		&			--	\\
RXJ1842.9-3532$^{2}$   & LMS & 154	$\pm$	7		&	T$^{34}$	            &	160	&	6.2	&	30.8	&			8.709			&			8.17			&			7.673			&			--	        		&			6.31E-03			&			--	        		&				--	        	    	&		$\leq$	9.10E+30		&			--	\\
DoAr44$^{2}$	       & LMS & 146	$\pm$	7		&	T$^{25}$	            &	30	&	2.9	&	29.2	&			8.246			&			7.61			&			6.794			&			--	        		&			1.00E-02			&			--	        		&				--	        	    	&		$\leq$	2.17E+31		&			--	\\
IM Lup$^{1}$	       & LMS & 158	$\pm$	8		&	MSp$^{36}$	            &	--  & 31.6	&	474.0	&			8.089			&			7.739			&			6.938			&			1.20E-03			&			--	        		&			--	        		&		$\leq$		4.01E+30			&			--	            	&			--	\\
V1247 Ori$^{1,2,22}$ & IMS 	& 398	$\pm$	25		&	1Sp$^{37}$	            &	--	& 39.8	&	1194.0	&			8.203			&			7.408			&			6.344			&			9.12E-04			&			8.32E-03			&			3.63E-03			&		$\leq$		1.73E+31			&		$\leq$	1.62E+32		&		$\leq$	5.07E+31	\\
DN Tau$^{1}$	    & LMS	& 128$\pm$	7			&	--	                &   --  &	12.8	&	384.0	&			8.342			&			8.015			&			7.716			&			1.91E-03			&			--	        		&			--	        		&		$\leq$		3.30E+30			&			--	            	&			--	\\
GO Tau$^{1}$	    & LMS	& 145$\pm$	7			&	--	                &	--	&	14.5	&	435.0	&			9.776			&			9.332			&			9.006			&			1.91E-02			&			--	        		&			--	        		&		$\leq$		1.12E+31			&			--	            	&			--	\\
DL Tau$^{1}$	    &	LMS	& 159$\pm$	8			&	--	                &	--	&	31.8	&	477.0	&			8.679			&			7.96			&			6.973			&			--	        		&			1.74E-03			&			--	        		&				--	        	    	&		$\leq$	3.26E+30		&			--	\\
MWC 480$^{1}$	    &	IMS	& 162$\pm$	12			&	--	                &	--	&	32.4	&	486.0	&			6.262			&			5.527			&			4.913			&			6.31E-04			&			--	        		&			--	        		&		$\leq$		1.18E+31			&			--	            	&			--	\\
CI Tau$^{1}$	    &	LMS	& 159$\pm$	8			&	FD, rings$^{38}$	&	--	&	31.8	&	477.0	&			8.431			&			7.793			&			6.775			&			2.75E-03			&			--	        		&			--	        		&		$\leq$		6.72E+30			&			--	            	&			--	\\
HD163296$^{23}$	    & IMS &	101	$\pm$	12			&	FD, rings$^{39}$	&	--	&	20.2	&	303.0	&			5.531			&			4.779			&			3.706			&			--	        		&			--	        		&			1.00E-03			&				--	        	    	&			--	            	&		$\leq$	1.03E+31	\\
HL Tau$^{24}$	    & LMS &	140	$\pm$				&	FD, rings$^{40}$	&	--	&	28.0	&	168.0	&			9.171			&			7.41			&			5.298			&			--	        		&			--	        		&			1.00E-03			&				--	        	    	&	  		--	            	&		$\leq$	4.52E+30	\\
DoAr 21$^{2}$       & LMS & 134	$\pm$	9			&	DD$^{41}$	        &	100	&	2.7	    &	26.8	&			6.862			&			6.227			&			5.783			&			--	        		&			1.91E-02			&			--	        		&				--	        	    	&		$\leq$	1.25E+32		&			--	\\
TYC 4496-780-1$^{1}$ & IMS	& 180$\pm$	12			&	--	                &	--	&	36.0	&	540.0	&			7.758			&			7.57			&			7.159			&			3.31E-03			&			--	        		&			--	        		&		$\leq$		1.93E+31			&			--          		&			--	\\
IRAS 04028+2948$^{1}$ & IMS	& 344$\pm$	13			&	--	                &	--	&	34.4	&	1032.0	&			9.472			&			8.831			&			7.722			&			2.75E-02			&			--	        		&			--	        		&		$\leq$		1.21E+32			&			--          		&			--	\\
V1075 Tau$^{1}$	    & LMS &	143	$\pm$	7			&	--	                &	--	&	14.3	&	429.0	&			9.056			&			8.85			&			8.733			&			3.02E-04			&			--	        		&			--	        		&		$\leq$		3.39E+29			&			--	            	&			--	\\
V1076 Tau$^{1}$	    & LMS &	151	$\pm$	7			&	--	                &	--	&	15.1	&	453.0	&			9.46			&			9.308			&			9.148			&			5.75E-04			&			--	        		&			--	        		&		$\leq$		4.92E+29			&			--	            	&			--	\\
V397 Aur$^{1}$	    & LMS &	149	$\pm$	18			&	--	                &	--	&	14.9	&	447.0	&			8.317			&			8.129			&			8.074			&			3.98E-03			&			--	        		&			--	        		&		$\leq$		9.46E+30			&			--	            	&			--	\\
V1207 Tau$^{1}$	    & LMS &	125	$\pm$	6			&	--	                &	--	&	12.5	&	375.0	&			8.960			&			8.802			&			8.734			&			2.09E-03			&			--	        		&			--	        		&		$\leq$		1.94E+30			&			--	            	&			--	\\
HIP 77545$^{1}$	    & IMS &	151	$\pm$	8			&	--	                &	--	&	30.2	&	453.0	&			7.996			&			7.895			&			7.829			&			3.31E-04			&			--	        		&			--	        		&		$\leq$		1.09E+30			&			--	            	&			--	\\
HIP 79462$^{1}$	    & IMS &	152	$\pm$	7			&	--	                &	--	&	30.4	&	456.0	&			7.429			&			7.294			&			7.261			&			9.12E-04			&			--	        		&			--	        		&		$\leq$		5.15E+30			&			--	            	&			--	\\
HIP 80088$^{1}$	    & IMS & 144	$\pm$	8			&	--	                &	--	&	28.8	&	432.0	&			7.904			&			7.785			&			7.728			&			8.32E-04			&			--	        		&			--	        		&		$\leq$		2.71E+30			&			--	            	&			--	\\
LkCa 19$^{1}$       & LMS & 160	$\pm$	13			&	DD$^{34}$	        &	190	&	16.0	&	480.0	&			8.318			&			8.148			&			8.058			&			2.75E-03			&			--	        		&			--	        		&		$\leq$		7.55E+30			&			--	            	&			--	
\enddata
\label{tab:t1}
\caption{Stars with protoplanetary disks imaged within 0$\arcsec$.25 with sufficient sensitivity to detect a 0.1~L$_{\sun}$ companion.}
\tablecomments{
[1]\citealt{Uyama2017},
[2]\citealt{Willson2016},
[3]\citealt{cugno2019},
[4]\citealt{quanz2013},
[5]\citealt{Sissa2018},
[6]\citealt{currie2017},
[7]\citealt{Follette2017},
[8]\citealt{Grady2013},
[9]\citealt{Reggiani2018k},
[10]\citealt{maire2017},
[11]\citealt{Gratton2019},
[12]\citealt{Ligi2018},
[13]\citealt{Mawet2017},
[14]\citealt{Christiaens2018},
[15]\citealt{Keppler2018},
[16]\citealt{haffert2019},
[17]\citealt{Kraus2012},
[18]\citealt{Sallum2015},
[19]\citealt{Cieza2013},
[20]\citealt{Canovas2018}
[21]\citealt{canovas2017},
[22]\citealt{Kraus2013},
[23]\citealt{Guidi2018},
[24]\citealt{Testi2015},
[25]\citealt{espaillat14}, 
[26]\citealt{Tang2017},
[27]\citealt{Akiyama2016},
[28]\citealt{Grady2001},
[29]\citealt{Follette2017},
[30]\citealt{Grady2013},
[31]\citealt{Muto2012},
[32]\citealt{Gratton2019},
[33]\citealt{Malfait1998},
[34]\citealt{vanderMarel2016},
[35]\citealt{Currie2011},
[36]\citealt{Avenhaus2018},
[37]\citealt{Dong2018b},
[38]\citealt{Clarke2018},
[39]\citealt{Zhang2016},
[40]\citealt{ALMA2015},
[41]\citealt{Jensen2009}}
\tablenotetext{a}{Transition disks are indicated by T, debris disks by DD, and full disks by FD. One-armed, two-armed, and multiarmed spirals are indicated by 1Sp, 2Sp, and MSp respectively. Disks with imaged ring structure are indicated by "rings".}
\end{deluxetable}

\clearpage
\end{turnpage}

\begin{deluxetable*}{lccccccccccc}
\tabletypesize{\scriptsize}
\tablehead{ \colhead{Star}	&	\colhead{Mass}	&	\colhead{dist}			
&\multicolumn{3}{c}{Stellar Magnitude}	
&	\multicolumn{3}{c}{Contrast}
&	\multicolumn{3}{c}{$\lambda$L$_{\lambda}$ (erg s$^{-1}$)}			\\
\colhead{}	&	\colhead{}	&	\colhead{pc}			
&	\colhead{H}	&	\colhead{K$\rm_s$}	&	\colhead{L}		
&	\colhead{$\Delta$ H}	&	\colhead{$\Delta$K$\rm_s$}	&	\colhead{$\Delta$L}	
&	\colhead{H}	&	\colhead{K$\rm_s$}	&	\colhead{L}	}
	
\startdata	
HD 142527$^{1}$ 	                &	    IMS	    &	    157$\pm$    1	    &   	5.715	    &	    4.980	    &   	4.280	    &		--	        	&		--	        	&		--	        	&		4.39E+32	        &		2.35E+32    	&		8.12E+31	\\
LkCa 15b$^{2, 3, 4}$	            &		LMS		&		159	$\pm$	8		&		8.600		&		8.163		&		7.49		&		5.25E-03		&		3.98E-03		&		1.00E-02		&			1.10E+31		&		6.16E+30    	&		7.72E+30	\\
LkCa 15c$^{4}$		                &	    LMS		&		159	$\pm$	8		&		8.600		&		8.163		&		7.49		&		5.25E-03		&		3.98E-03		&		1.00E-02		&		9.70E+30	        &		6.16E+30		&	--	\\
PDS 70b$^{5, 6}$	            	&		LMS		&		113	$\pm$	5		&		8.823		&		8.542		&		8.026		&		--	        	&		6.31E-04		&		1.74E-03		&			--	        	&		3.51E+29	    &		4.18E+29	\\
PDS 70c$^{6}$	                   	&		LMS		&		113	$\pm$	5		&		8.823		&		8.542		&		8.026		&		--	        	&		3.02E-04		&		2.29E-03		&			--	        	&		1.68E+29	    &		5.46E+29	
\enddata
\label{tab:t2}
\caption{Stars with accreting companions imaged.}
\tablecomments{[1]\citealt{Christiaens2018},
[2]\citealt{Uyama2017},
[3]\citealt{Kraus2012},
[4]\citealt{Sallum2015},
[5]\citealt{Keppler2018},
[6]\citealt{haffert2019}} 

\end{deluxetable*}

\begin{deluxetable*}{lcccc}

\tabletypesize{\scriptsize}

\tablehead{ \colhead{Star}	& \colhead{a}   & \colhead{M}   & \colhead{M\.M} & \colhead{$\rm L/L_{\sun}$} \\
           \colhead{}       & \colhead{(au)}  & \colhead{($\rm M_J$)} & \colhead{($\rm M^2_J~yr^{-1}$)} & \colhead{}}
\startdata	
HD 142527B$^1$	&	$22^{+19}_{-11}$ &			$270^{+170}_{-150}$         & $10^{-2}$ [4]         & 0.6\\
LkCa 15b$^2$	&	$14.7\pm2.1$	&			$\leq10$	                & $10^{-5}$             & $\sim10^{-3}$\\
LkCa 15c$^2$	&	$18.6\pm2.5$	&			$\leq10$	                & $10^{-5}$             & $\sim10^{-3}$\\
PDS 70b$^3$	    &	$20.6\pm1.2$	&			4--17	                    & $0.3-9\times10^{-7}$  & $\sim10^{-4}$\\
PDS 70c$^3$	    &	$34.5\pm2.0$	&			4--12	                    & $0.15-6\times10^{-7}$ & $\sim10^{-4}$
\enddata
\caption{Companion data}
\tablecomments{[1]\citealt{Christiaens2018}, [2] \citealt{Sallum2015}, [3] \citealt{haffert2019}, [4] M\.{M} is estimated by assuming that luminosity of the companion object is dominated by the accretion luminosity and that the radius of the companion is 1.5$\rm R_J$.}
\end{deluxetable*}

\begin{figure*}
\begin{center}
\includegraphics[scale=0.9,angle=0,trim={0.5in 1in .5in 2in},clip]{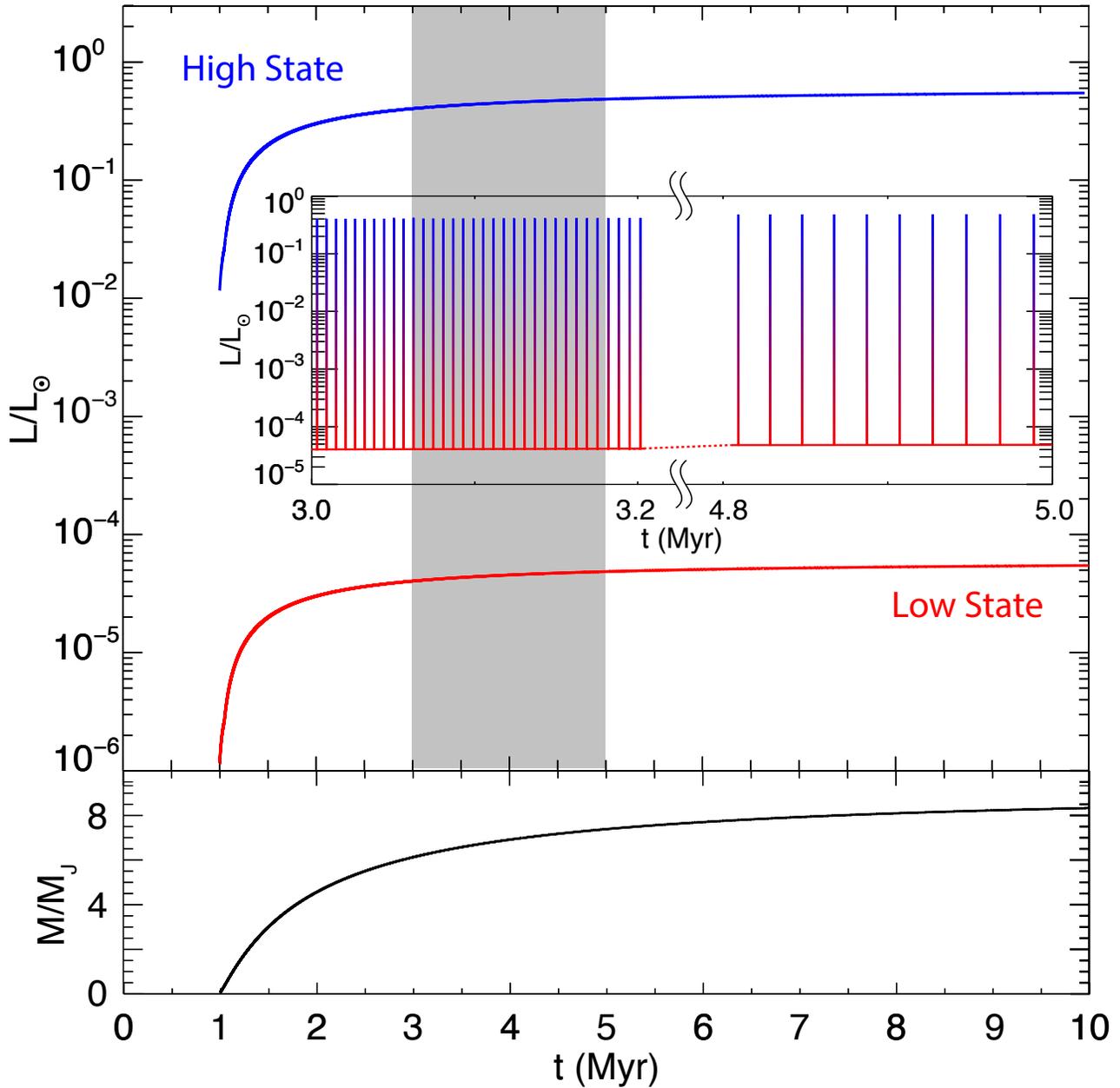}
\caption{Accretion luminosity (upper panel) and planetary mass (lower panel) as a function of time. The accretion luminosity increases with planetary mass and alternates between quiescence (red curve) and
outburst (blue curve). The inset shows an enlarged version of the shaded region of the plot. When the planet mass is low and the circumstellar disk accretion rate is high, outbursts are more frequent as the circumplanetary disk grows in mass to 0.1~M$\rm_p$ more frequently. However the potential well is not as deep and the accreted mass is lower, so the luminosity of the accretion is lower. As the planet grows in mass and the circumstellar disk accretion rate declines, the outbursts become less frequent but more intense. In the model shown, the planet mass grows to $\sim$8~$M_J$ in 5~Myr and continues to grow until the circumstellar disk ceases to feed the circumplanetary disk and the circumplanetary disk empties all of its mass onto the central object. \label{fig:1}}
\end{center}
\end{figure*}

\begin{figure*}
\begin{center}

\includegraphics[scale=0.7,angle=-90,trim={0in 0in 0in 0in},clip]{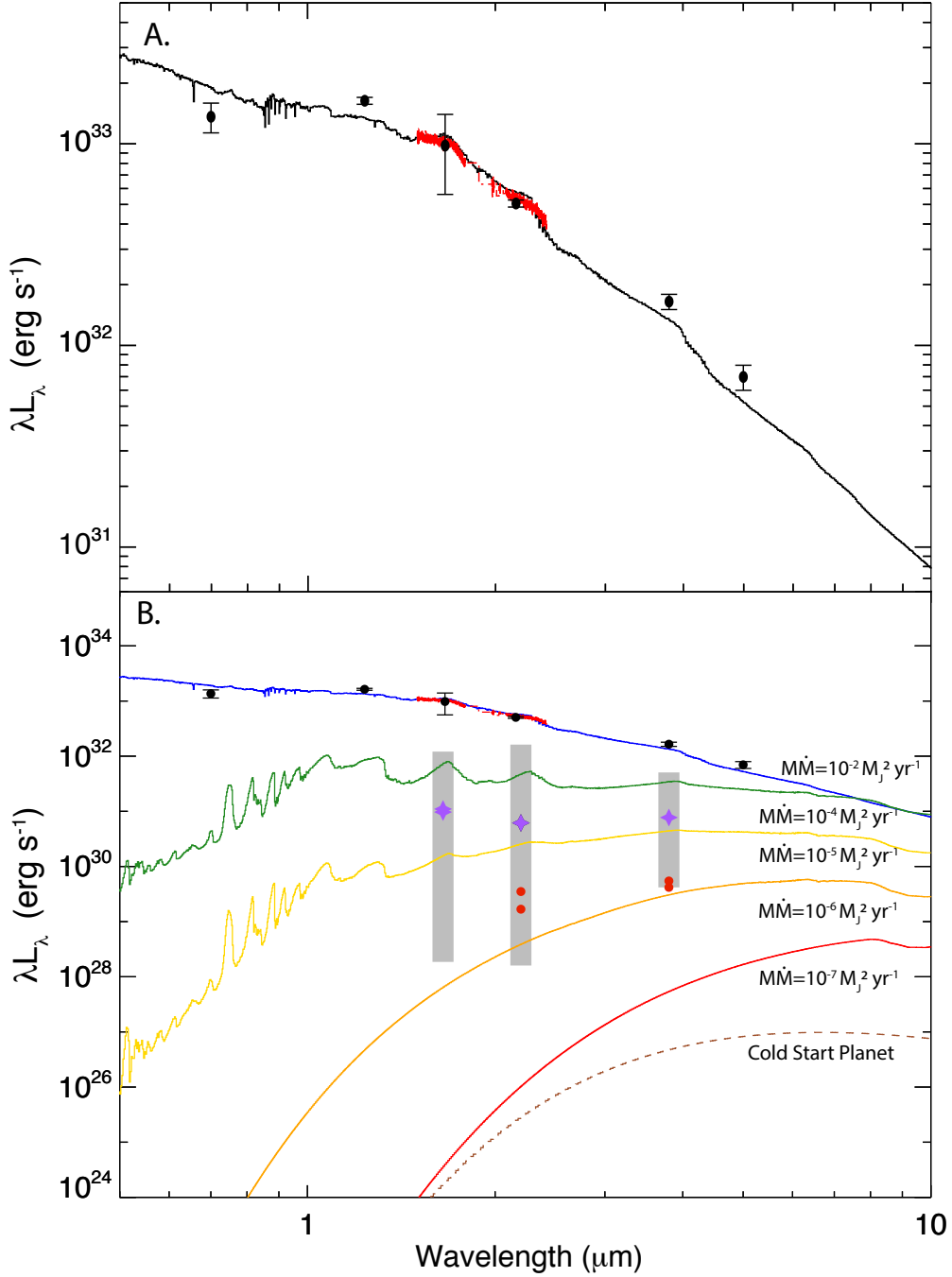}

%\label{fig:3b}
%\end{subfigure}

\caption{Photometry of 
the companion orbiting HD142527 and models of active circumplanetary disks.  
The companion photometry is corrected for circumplanetary 
disk inclination and reddening following \citet{Zhu2015} assuming an inclination of $70\degr$, $A_J=1.6$ and $\alpha=1.7$. With these parameters, the photometry of the companion is well fit by a model of an active disk with $M\dot{M}=10^{-2}~M_J^2~yr^{-1}$, $R_{in}=1.5~R_J$, and $R_{out}=15~R_J$ (Panel A).   In Panel B, we replot the
companion  photometry 
as well as the photometry of the candidate planets orbiting
LkCa~15 (purple stars) and PDS~70 (red circles). The range of observational limits from the literature are shown as vertical gray bars. We also plot the SED of an accreting circumplanetary disk for four additional accretion rates as well
as the spectrum of a 550~K blackbody representing a cold-start 1~Myr old 1~M$\rm_J$ planet. 
Previous observations are not sensitive enough to have detected a gas giant planet with a quiescent disk ($M\dot{M} < 10^{-6}~M_J^2~yr^{-1}$). \label{fig:3}}

\end{center}

\end{figure*}

\end{document}